\documentclass[aps,prb,twocolumn,floats]{revtex4}

\usepackage{amssymb}
\usepackage{amsmath}
\usepackage{graphicx}
\usepackage{bm}
\usepackage{mdframed}
\usepackage[pangram]{blindtext}

\begin{document}

\bibliographystyle{prsty}
\author{H\'{e}ctor Ochoa}
\affiliation{Department of Physics, Columbia University, New York, NY 10027, USA}
%\date{\today}

\begin{abstract}
In twisted bilayer graphene, long-wavelength lattice fluctuations on the scale of the moir\'e period are dominated by phason modes, i.e., acoustic branches of the incommensurate lattice resulting from coherent superpositions of optical phonons. In the limit of small twist angles, these modes describe the sliding motion of stacking domain walls separating regions of partial commensuration. The resulting soliton network is a soft elastic manifold, whose reduced rigidity arises from the competition between intralayer (elastic) and interlayer (adhesion) forces governing lattice relaxation. Shear deformations of the beating pattern dominate the electron-phason coupling to the leading order in $t_{\perp}/t$, the ratio between interlayer and intralayer hopping parameters. This coupling lifts the layer degeneracy of the Dirac cones at the corners of the moir\'e Brillouin zone, which could explain the observed 4-fold (instead of 8-fold) Landau level degeneracy. Electron-phason scattering gives rise to a linear-in-temperature contribution to the resistivity that increases with decreasing twist angle due to the reduction of the stiffness of the soliton network. This contribution alone, however, seems to be insufficient to explain the huge enhancement of the resistivity of the normal state close to the magic angle.
\end{abstract}
%\pacs{}

\title{Moir\'e-pattern fluctuations and electron-phason coupling in twisted bilayer graphene}

\maketitle

\section{Introduction}

Two graphene layers rotated with respect to each other an incommensurate angle $\theta$ form a quasi-periodic structure known as moir\'e pattern.\cite{portu} At a \textit{magic angle} $\theta\sim 1^{\textrm{o}}$, quantum interference of electrons in the associated superlattice potential gives rise to narrow low-energy bands,\cite{macdonald} setting the stage for strongly correlated phenomena.\cite{jarillo1,jarillo2} In addition to what appears to be Mott insulating states at half-filling of the lowest flat bands\cite{jarillo1} and the onset of superconductivity under doping\cite{jarillo2} or hydrostatic pressure,\cite{columbia} new many-body insulating states have been reported for different filling factors, some of them with apparent topological character.\cite{stanford,efetov,UCSB}

Qualitative differences in phenomenology from one device to another suggest that we should envision the moir\'e pattern not as a rigid potential landscape but as a spatially inhomogeneous and (as I will argue here) most likely fluctuating one. Structural inhomogeneities around the magic angle are revealed by variations in the electronic densities of full-filled (single-particle) insulating states, along with evidences from quantum interference measurements of the formation of insulating islands within the superconducting state.\cite{jarillo2,columbia} For smaller twist angles, spatial variations of the moir\'e period have been directly visualized by transmission electron \cite{TEM} and scanning tunnel microscopies\cite{LeRoy} as well as near-field optical techniques,\cite{Basov} showing the formation of regions of partial commensuration separated by stacking domain walls.\cite{domain_walls_exp,domain_walls_th} These structural differences can induce/favor different symmetry-broken states. The interaction with the encapsulating boron nitride in transport devices can break the sublattice symmetry (usually referred as $\mathcal{C}_2\mathcal{T}$ symmetry in the literature\cite{symmetry1,symmetry2}), opening a gap in the low-energy bands, which acquire nonzero valley-Chern numbers. This state could serve as a precursor for the formation of a (quantum) anomalous Hall ferromagnet at odd fillings as the bands become spin-valley polarized due to electron correlations.\cite{stanford,Zaletel,Zhang_etal} Mean-field calculations suggest that this symmetry can also be spontaneously broken,\cite{Xie_MacDonald} which has been invoked to explain some features in the magnetotransport of the most homogeneous samples at one-quarter filling of the hole band.\cite{efetov} The electronic spectrum is also sensitive to the dielectric environment due to electrostatic effects.\cite{Paco_PNAS} In open samples, suitable for local probes, tunneling spectroscopy has unveiled the emergence of charge ordering with broken 3-fold symmetry as the Fermi level varies.\cite{STM1,STM2,STM3} 

Another striking observation is the remarkably large, linear in temperature ($T$) resistance of the normal state.\cite{jarillo3,columbia_phonons} Here there appear to be subtle difference between samples too: While in the devices of Cao \textit{et al}.\cite{jarillo3} a linear-$T$ resistivity is only apparent for fractional fillings (more prominently at half-filling), suggesting a connection with the correlated state at lower temperatures, the devices of Polshyn \textit{et al}.\cite{columbia_phonons} show qualitatively the same behavior (albeit non-monotonic) for a broader range of carrier densities, with increasing values of the resistivity as the twist angle decreases. This latter behavior has been attributed to electron-phonon scattering.\cite{phonons_transport1,phonons_transport2,phonons_transport3} These studies focus on the original acoustic modes of graphene, but as noticed in Ref.~\onlinecite{phonons_Bernevig}, relative displacements of the two layers have a stronger impact on the moir\'e interference pattern and, potentially, on the electronic structure. Here I analyze the case with the account of interlayer interaction forces, which play a fundamental role in the energetics of these modes.%, determining, for example, the sensibility of the moir\'e pattern to disorder.

The structure of the article is as follows. I will discuss first in Sec.~\ref{sec:mechanics} the emergence of new collective modes, \textit{phasons}, associated with the broken translational symmetry of the incommensurate lattice, or more accurately, the invariance of the system under relative translations of the two layers. These modes define two acoustic branches in the spectrum of small oscillations, which is determined by the competition of the two relevant length scales in the problem: the moir\'e period, and a length scale related to the curvature of the adhesion potential that define the characteristic width of stacking textures (solitons) connecting degenerate adhesion-energy minima. Already for angles $\theta\lesssim 3^{\textrm{o}}$, phonon softening is very pronounced, marking the instability of the system towards the formation of a soliton network. Phason dynamics is then governed by the effective elasticity of these objects. In Sec.~\ref{sec:e-ph}, I will analyze how phasons couple to electrons. Transverse modes (i.e., shear and rotational deformations of the interference pattern) have a stronger impact on the low-energy spectrum and, in particular, can explain the reduced Landau level degeneracy observed in magnetotransport.\cite{jarillo1,jarillo2,columbia} Although the coupling considered here is intrinsically weak (proportional to the ratio between interlayer and intralayer hopping parameters), scattering by large (due to the reduced rigidity of stacking solitons) phason fluctuations limit the electron mobilities, leading to a linear-$T$ resistivity that increases with decreasing twist angle. However, this contribution seems to be insufficient to explain the dramatic increase of the resistivity around the magic angle reported in the experiments. I will finally discuss other possible scenarios in Sec.~\ref{sec:discussion}. Details of the calculations are saved for the appendices.

\section{Phasons in moir\'e-patterned bilayer graphene}

\label{sec:mechanics}

\begin{figure}[t!]
\begin{center}
%\hspace{-0.4cm}
\includegraphics[width=\columnwidth]{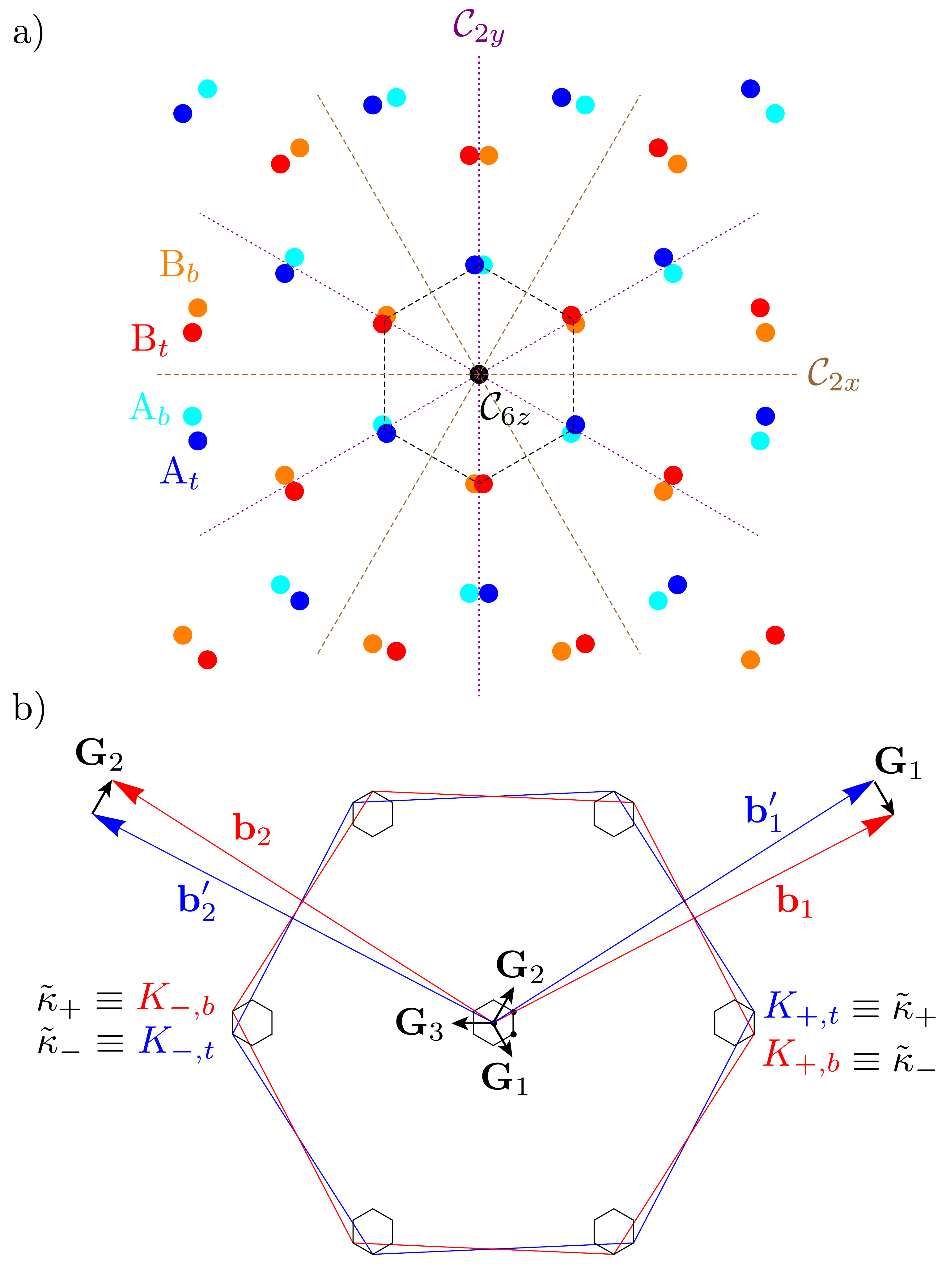}
\caption{a) Microscopic lattice (no relaxation) around a beating pattern maximum (local AA stacking) for a twist angle of $\theta=5^{\textrm{o}}$. Dashed lines highlight the hexagonal ($D_6$) symmetry of the continuum model, with a 6-fold rotation axis along the common center and 2-fold rotation axis within the plane. b) Superimposed Brillouin zones of top (in blue) and bottom (in red) layers. In commensurate approximants, valleys $K_{\tau,\mu}$ lie at the two inequivalent corners of the moir\'e Brillouin zone, $\tilde{\kappa}_{\eta}$; hereafter $\tau=\pm 1$ labels the valley, $\mu=\pm1 \left(t/b\right)$ labels the top ($t$) and bottom ($b$) layers, and $\eta=\tau\times\mu$ labels the corresponding $\tilde{\kappa}_{\eta}$ point in type-I or sublattice-exchange odd commensurate structures,\cite{Mele0,portu2,symmetry2} which are dense in the limit of small twist angles. %The insets illustrate the splitting of the Dirac mini-bands under a shear or local twist distortion of the moir\'e pattern; dashed lines correspond to valley $\tau=+1$, straight lines to valley $\tau=-1$, blue color to top layer $\mu=+1\,(t)$, and red color to bottom layer $\mu=-1\,(b)$.
}
%\vspace{-0.5cm} 
\label{fig:symmetry}
\end{center}
\end{figure}

Let me first introduce the geometry of the moir\'e superlattice. In the absence of lattice relaxation, atomic positions in the bottom and top layers are spanned by primitive vectors $\mathbf{a}_{1,2}$ and $\mathbf{a}_{1,2}'=R(\theta)\,\mathbf{a}_{1,2}$ of a triangular Bravais lattice, where $R(\theta)$ is a SO(2) matrix describing the relative rotation. For concreteness, I am going to consider the situation depicted in Fig.~\ref{fig:symmetry}, in which carbon atoms sit on top of each other (AA stacking) prior to the relative rotation. This choice is arbitrary but ultimately inconsequential in the limit of small (generically incommensurate) twist angles, for in that case the system explores all possible stacking configurations. For future reference, lateral positions will be referred to the rotation axis; the coordinate system is such that the $x$ and $y$ axes lie along the two-fold symmetry axes highlighted in Fig.~\ref{fig:symmetry}~a. Positive twist angles will correspond to anti-clockwise rotations of the top layer. 

The moir\'e superlattice is defined by a beating pattern resulting from the periodicity of the individual triangular lattices, i.e., the Fourier components of the atomic density $\rho(\mathbf{r})$ are peaked at vectors $\mathbf{G}$ of the incommensurate reciprocal lattice,\begin{align}
\label{eq:mass_density}
\rho\left(\mathbf{r}\right)\approx\sum_{\left\{\mathbf{G}\right\}}\rho_{\mathbf{G}}\,e^{i\mathbf{G}\cdot\mathbf{r}}=\sum_{\left\{\mathbf{G}\right\}}\left|\rho_{\mathbf{G}}\right|\,e^{-i\phi_{\mathbf{G}}+i\mathbf{G}\cdot\mathbf{r}},
\end{align}
where in the last expression I have separated the Fourier components in modulus and phase (note that $\phi_{-\mathbf{G}}=-\phi_{\mathbf{G}}$, so $\rho(\mathbf{r})$ is real). The set $\{\mathbf{G}\}$ corresponds to the lattice spanned by primitive vectors $\mathbf{G}_{1,2}=\mathbf{b}_{1,2}-\mathbf{b}_{1,2}'$, where $\mathbf{b}_{1,2}$ and $\mathbf{b}_{1,2}'=R(\theta)\, \mathbf{b}_{1,2}$ are primitive vectors of the bottom and top reciprocal lattices, respectively (see Fig.~\ref{fig:symmetry}~b). The moir\'e superlattice is just the dual to $\{\mathbf{G}\}$, spanned by primitive vectors\cite{foot1}\begin{align}
\label{eq:geometry}
\mathbf{R}_{1,2}=\left[1-R^{-1}\left(\theta\right)\right]^{-1}\, \mathbf{a}_{1,2}.
\end{align}
It is also convenient to introduce the function $\boldsymbol{\Delta}(\mathbf{r})\equiv\mathbf{r}-R^{-1}(\theta)\,\mathbf{r}$, which measures the distance between a given point $\mathbf{r}$ in the top layer with respect to its original position in the bottom layer before the twist (assuming yet no lattice relaxation). Note that at the maxima (local AA stacking) of the interference pattern $\boldsymbol{\Delta}$ coincides with a lattice vector of the graphene Bravais lattice, $\boldsymbol{\Delta}(n\,\mathbf{R}_{1}+m\,\mathbf{R}_{2})=n\,\mathbf{a}_{1}+m\,\mathbf{a}_{2}$, with $n$, $m$ integers. The beating pattern maxima are separated by a distance $L_M=|\mathbf{R}_{1,2}|=\sqrt{3}a/(2\sin\theta/2)$, defining the moir\'e superlattice period.

So far, interlayer forces have not appeared in our discussion. Lattice relaxation of structures with a nominal twist angle ($|\rho_{\mathbf{G}}|\neq0$) could be generically described by a Landau-like free-energy expansion in powers of the beating-pattern density which, as inferred from the previous discussion, gives us an idea of the degree of overlap in the lateral position of the atoms in both layers (and therefore of the interlayer coupling). Minimization of this functional would lead to constraints of the form $\sum_{i=1...n}\phi_{\mathbf{G}_i}=\gamma$, with $\sum_{i=1...n}\mathbf{G}_i=0$, coming from terms in the $n$th power of $\rho(\mathbf{r})$. The reader should note that such \textit{relaxed structures} are only metastable states that, in real devices, are likely to be stabilized by the unavoidable tensions generated during the fabrication process. In fact, it is likely that the samples present patches with different twist angles, i.e., different period of the moir\'e patterning. Here we are concerned only about long-wavelength fluctuations (smooth on the scale of the moir\'e period $L_M$) around a local minimum.

The previous constraints defining the configuration of minimum energy leave a certain number $N$ of the phases $\phi_{\mathbf{G}_i}$ unspecified; in other words, variations of $N$ phases generate physically distinct but energetically equivalent quasi-periodic structures and, thus, represent soft modes of the system. In this case, the number of soft modes is $N=2$, for $\{\mathbf{G}\}$ is a two-dimensional Bravais lattice. Consider, for example, a beating pattern truncated to the first six Fourier harmonics (first star), which set the six-fold symmetry and the single independent length scale of the moir\'e superlattice: $\rho(\mathbf{r})\approx \rho_0+\rho_1\sum_{i=1,2,3}\cos(\mathbf{G}_i\cdot\mathbf{r}-\phi_{\mathbf{G}_i})$, with $\mathbf{G}_3\equiv-\mathbf{G}_1-\mathbf{G}_2$. A generic cubic term in the phenomenological free energy fix one of the three phases. A convenient parametrization is then of the form\begin{align}
\label{eq:parametrization}
\phi_{\mathbf{G}_i}=\mathbf{\tilde{u}}\cdot\mathbf{G}_{i}+\frac{\gamma}{3},
\end{align}
where $\mathbf{\tilde{u}}$ is a two-dimensional vector describing the \textit{phason} modes of the incommensurate lattice.\cite{{Levine_etal}} Note that changes in $\mathbf{\tilde{u}}$ translate rigidly the beating pattern, while an arbitrary change in $\gamma$ (which is not a soft mode) distorts the pattern. In a totally incommensurate or \textit{floating} state, i.e., when lattice relaxation is ignored, $\mathbf{\tilde{u}}$ can be identified straightforwardly with a relative displacement between layers, $\mathbf{u}$. Note that in that case $\boldsymbol{\Delta}(\mathbf{r})\rightarrow \mathbf{r}-R^{-1}(\theta)(\mathbf{r}-\mathbf{u})=\boldsymbol{\Delta}(\mathbf{r})+R^{-1}(\theta)\,\mathbf{u}$, therefore, the maxima of the beating pattern are translated a distance\cite{foot2} \begin{align}
\label{eq:distances}
\mathbf{\tilde{u}} & =\left[1-R\left(\theta\right)\right]^{-1}\,\mathbf{u}=\left(\frac{1}{2}+\frac{1}{2}\cot\frac{\theta}{2}\,\,\mathbf{\hat{z}}\times\right)\mathbf{u}.
%\\
%& =\left[\begin{array}{cc}
%\frac{1}{2} & -\frac{1}{2}\cot\frac{\theta}{2} \\
%\frac{1}{2}\cot\frac{\theta}{2} & \frac{1}{2}
%\end{array}\right]\left(\begin{array}{c}
%u_x\\
%u_y
%\end{array}\right),
%\nonumber
\end{align}
%In this limit, the energetics of phason can be described by the elasticity of individual layers (CITAS), which is a good approximation in the limit of small moir\'e cells or large twist angles, or alternatively, weak interlayer coupling (I will give a precise definition of this regime in a few lines).

\begin{figure*}[t!]
\begin{center}
%\hspace{-0.4cm}
\includegraphics[width=\textwidth]{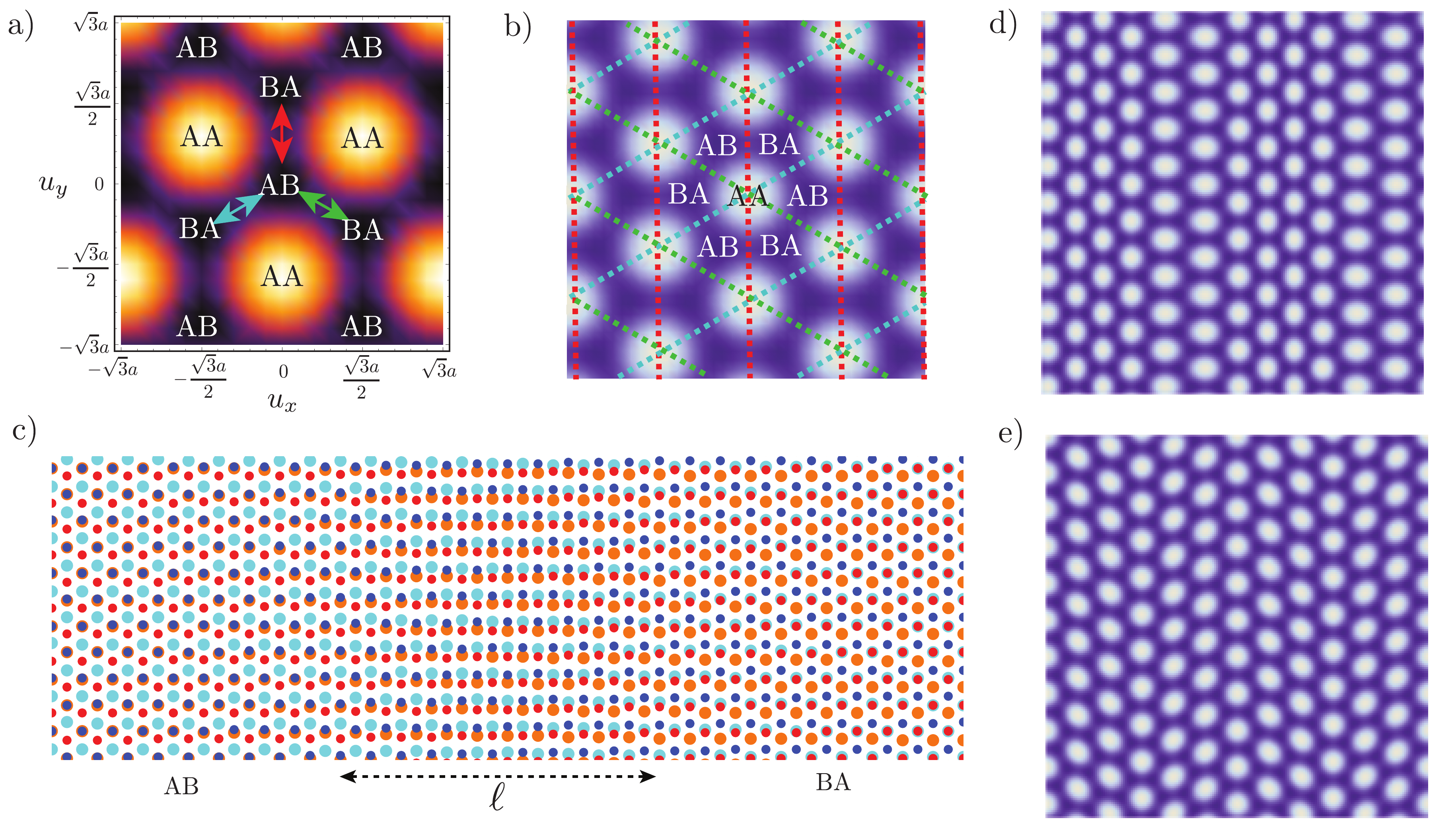}
\caption{a) Adhesion-energy landscape deduced from Eq.~\eqref{eq:potential}. Arrows represent stacking textures connecting degenerate minima (AB and BA Bernal stackings). b) Generic beating pattern (Eq.~\ref{eq:mass_density}) formed by two floating layers. When $\ell$ is ostensibly smaller than $L_M$, sharper stacking textures are formed, corresponding to domain walls between regions of partial commensuration. The domain walls are represented as dashed lines in the same colors as their representation in configurational space, panel~a. c) Sine-Gordon shear soliton connecting AB and BA stacking regions. Colored dots represent atomic positions as in Fig.~\ref{fig:symmetry}. d) Longitudinal phason distortion of the beating pattern with wavelength $\lambda=3\sqrt{3}L_M$. e) Transverse phason with the same wavelength.}
%\vspace{-0.5cm} 
\label{fig:fig2}
\end{center}
\end{figure*}

More generically, phason modes correspond to coherent superpositions of optical phonons with momenta separated by a superlattice vector $\mathbf{G}$. When lattice relaxation is taken into account, these coherent superpositions describe more complex atomic rearrangements than a simple relative displacement of the two layers. The coordinate $\mathbf{\tilde{u}}$ can be identified then with the sliding motion of domain walls separating regions of partial commensuration (alternating AB and energetically equivalent BA Bernal stackings, where atoms of different sublattices lie on top of each other).%, as sketched in Fig.~?. %The construction here is similar to the formation of incommensurate structures of adsorbed atomic layers on graphite, where two-dimensional soliton networks can be formed from simpler one-dimensional solutions \cite{Pokrovsky_Tapalov} imposing simple rules\cite{Villain} in order to minimize, for example, the number of soliton crossings.\cite{foot3}

\subsection{Soliton network}

The problem of lattice relaxation can be addressed in the framework of a two-dimensional version of the Frenkel-Kontorova model.\cite{rusos,Mele,Koshino} Neglecting entropic terms in the free energy, the problem reduces to solve the following equation for $\mathbf{u}(\mathbf{r})$, understood now as a field in the continuum describing smooth (in the scale of the interatomic distance $a\approx 1.42$ \AA) relative displacements of the layers:\begin{align}
\label{eq:FK}
\frac{\lambda+\mu}{2}\,\boldsymbol{\nabla}\left(\boldsymbol{\nabla}\cdot\mathbf{u}\right)+\frac{\mu}{2}\,\nabla^2\mathbf{u}=\frac{\partial}{\partial\mathbf{u}}\mathcal{V}_{\textrm{ad}}\left[\mathbf{r},\mathbf{u}\left(\mathbf{r}\right)\right].
\end{align}
Here $\mu\approx 3\lambda\approx 9$ eV/\AA$^2$ are the Lam\'e coefficients of graphene;\cite{Lame} I am disregarding corrugations provided that the bending energy is inconsequential on length scales longer than $\sqrt{\kappa/(\lambda+2\mu)}\sim a$, where $\kappa\approx 1$ eV is the bending rigidity.

The right-hand side of Eq.~\eqref{eq:FK} describes variations in configurational space (stackings) of the adhesion potential (here with units of energy density) between layers. The notation emphasizes the different periodicities in real and configurational spaces: while the dependence on $\mathbf{r}$ is modulated on the scale of the moir\'e pattern, i.e., it admits a Fourier expansion in $\{\mathbf{G}\}$, the dependence on stacking configurations changes on the atomic unit cell. A first-star expansion of the latter compatible with six-fold symmetry reduces to\begin{align}
\label{eq:potential}
\mathcal{V}_{\textrm{ad}}\left[\mathbf{u}\left(\mathbf{r}\right)\right]=V\sum_{i=1}^3\left\{\frac{1}{2}+\cos\left[\mathbf{b}_i\cdot\left(\mathbf{u}+\boldsymbol{\Delta}_0\right)\right]\right\},
\end{align}
where $\boldsymbol{\Delta}_0$ is the displacement of the top layer with respect to the bottom layer in Bernal (AB) stacking, which sets the reference in energies. The phenomenological parameter $V\approx 90$ meV/nm$^2$ measures the energy difference between Bernal and AA stackings.\cite{Carr_etal} The adhesion-energy landscape deduced from this model is shown in Fig.~\ref{fig:fig2}~a. In this approximation, the modulation in real space follows from Eq.~\eqref{eq:potential} just by noting that when layers are rotated, the separation between unit cells depends on the position $\mathbf{r}$ via the substitution $\boldsymbol{\Delta}_0\rightarrow\boldsymbol{\Delta}(\mathbf{r})$, and $\mathbf{b}_i\cdot\mathbf{\Delta}\left(\mathbf{r}\right)=\mathbf{G}_i\cdot\mathbf{r}$ from the previous definitions.

In addition to the periodicity of the moir\'e pattern, there is another length scale encrypted in Eq.~\eqref{eq:FK} related to the curvature of the adhesion potential,\begin{align}
\label{eq:length}
\ell=\sqrt{\frac{\mu}{2\,\frac{\partial^2\mathcal{V}_{\textrm{ad}}}{\partial u^2}|_{\textrm{AB}}}}=\frac{a}{\pi}\sqrt{\frac{\mu}{2V}}\approx 3.2\,\,\text{nm}.
\end{align}
This length scale characterizes the spatial extension of stacking textures connecting equivalent minima (AB and BA stackings). The competition between their cost in elastic energy and the adhesion energy of large incommensurate areas governs the degree of lattice relaxation. When $L_M$ and $\ell$ are comparable, lattice relaxation is negligible and the beating pattern is well approximated by two floating layers. However, when $L_M$ is larger than $\ell$, it is energetically cheaper for the system to form regions of partial commensuration separated by domain walls of characteristic width $\ell$, where the cost in elastic and adhesion energies is concentrated.

The competition between these two length scales is reflected in the the spectrum of small oscillations around a local minimum, see Appendix~\ref{sec:A}. The harmonic expansion of the adhesion potential around a floating state only couples modes with momentum mismatch in the first star, originally separated by a frequency of the order of $\omega_M=4\pi c/(\sqrt{3}L_M)$, where $c=\sqrt{\mu/\rho_0}\approx14$ Km/s is the sound velocity of transverse phonons in graphene. When $L_M\sim\ell$, the strength of these harmonics is negligible with respect to $\omega_M$, and the spectrum resembles the acoustic phonons of single-layer graphene folded into the moir\'e Brillouin zone. When $L_M$ is ostensibly larger than $\ell$, this simple model predicts a strong softening of the two acoustic branches, indicating that the floating state is unstable and lattice relaxation is not longer negligible. Sharper stacking textures are formed,\cite{Koshino} giving rise to more harmonics in the expansion of the adhesion potential that have to be included in a new calculation of the spectrum of oscillations. Already in this regime, the simple identification in Eq.~\eqref{eq:distances} breaks down and the energetics of phasons are no longer described by the elasticity of individual layers.

The first step is to determine the stacking texture. In order to construct approximated long-wavelength density profiles, let me consider first one-dimensional solutions of Eq.~\eqref{eq:FK} of the form $\mathbf{u}(\mathbf{r})=u(\varrho)\,\mathbf{\hat{u}}$, where the spatial dependence (along unit vector $\boldsymbol{\hat{\varrho}}$) is not necessarily collinear with lattice relaxation. %The following construction is analogous to models in incommensurate structures of adsorbed atomic layers on graphite, where heuristic models for two-dimensional networks based on simpler one-dimensional solutions \cite{Pokrovsky_Tapalov} imposing simple rules in order to minimize, for example, the number of soliton crossings.\cite{Villain}
A simple inspection of Eq.~\eqref{eq:potential} (plugging the one-dimensional ansatz and projecting over $\mathbf{\hat{u}}$ and $\mathbf{\hat{z}}\times\mathbf{\hat{u}}$) shows that only \textit{tensile} ($\mathbf{\hat{u}}\parallel\boldsymbol{\hat{\varrho}}$) or \textit{shear} ($\mathbf{\hat{u}}\perp\boldsymbol{\hat{\varrho}}$) solutions exist in this approximation.\cite{foot3} I am going to focus on the latter, which are less energetic; this assumption agrees with numerical calculations\cite{Koshino,Carr_etal} (note also that the softening is more pronounced for the transverse oscillation mode) and is obviously favored by the orientation of the moir\'e superlattice with respect to the graphene lattice in the limit of small twist angles. As implied by the energy landscape in Figs.~\ref{fig:fig2}~a, the relaxation is more likely to occur along an armchair direction. Let me focus first on the case $\mathbf{\hat{u}}_3\approx\mathbf{\hat{y}}$, $\boldsymbol{\hat{\varrho}}_3\approx\mathbf{\hat{x}}$ (solitons marked in red in the figure); the other two solutions with the same energy follow from rotating the axes $120^{\textrm{o}}$. The notation emphasizes that $\mathbf{G}_i\approx -\frac{2\pi}{L}\boldsymbol{\hat{\varrho}}_i$ for small twist angles, with $L=|\mathbf{\hat{x}}\cdot\mathbf{R}_{1,2}|\approx \sqrt{3}L_M/2$. 

In the asymptotic limit $L_M\gg \ell$, we can neglect the modulation on the scale of the moir\'e (first argument in the adhesion potential) and write $\mu\, u''=(V\pi/a)\sin(2\pi u/a)$, whose general solution is a train of domain walls separated a certain fixed length. The latter should be determined from energetic considerations,\cite{book} but in our heuristic construction we can just identify this length scale with $L$. In this limit, the domain-wall profile is well described by a sine-Gordon soliton,
\begin{align}
\label{eq:domain_wall}
u_3\left(x\right)\approx\frac{2a}{\pi}\arctan\left(e^{\frac{x-x_3}{\ell}}\right),
\end{align}
as numerical calculations confirm.\cite{Koshino} The corresponding stacking texture is represented in Fig.~\ref{fig:fig2}~c. Associated with the domain wall, there is a tension $\sigma=a\sqrt{2\mu V}/\pi$ characterizing its energy cost. This energy does not depend on the soliton center, $x_3$: while the separation between domain walls is set by the moir\'e periodicity, rigid displacements of the soliton solution does not cost energy.

The density wave associated with the one-dimensional train of domain walls is of the form
\begin{align}
\label{eq:density_wave}
\rho_{\textrm{1D}}\left(\mathbf{r}\right) & =\sum_{n=0}^{\infty}\left|\rho_n\right|\cos\left[\frac{2\pi n}{L}\left(x-x_3\right)\right]\\
& \approx \sum_{n=0}^{\infty}\left|\rho_n\right|\cos\left[n\left(\boldsymbol{\hat{\varrho}}_3\cdot\mathbf{G}_3\right)\left(x-x_3\right)\right],
\nonumber
\end{align}
where the Fourier components $|\rho_n|$ can be estimated from the approximate profile in Eq.~\eqref{eq:domain_wall}. The two-dimensional beating pattern can be approximated then by the superposition of three density waves like Eq.~\eqref{eq:density_wave} with director vectors rotated $120^{\textrm{o}}$. At \textit{rigid} soliton crossings, the system explores all possible stacking configurations. Adhesion forces will relax the structure, but some areas will remain pinned to saddle points of the potential, including AA stackings, introducing a large free-energy cost. Therefore, the configuration of minimum energy is such that the number of soliton crossings is minimized, and consequently, from the three phases related to the soliton centers, $x_{1,2,3}$, only two are really independent. In the parametrization of Eq.~\eqref{eq:parametrization}, the soliton centers are related to the phason field as\begin{align}
x_i\approx\boldsymbol{\hat{\varrho}}_i\cdot\mathbf{\tilde{u}}-\frac{\gamma L}{6\pi}.
\end{align}
Minimizing the number of soliton crossings corresponds to the condition $x_0+x_1+x_2\approx0$ (modulo $L/3$), or equivalently, $\gamma$ must be an integer of $2\pi$.\cite{foot4} The resulting beating pattern can be envisioned as a triangular lattice of AA-stacked regions connected by sine-Gordon domain walls, as illustrated in Fig.~\ref{fig:fig2}~b.

\subsection{Generalized elasticity}

Provided that $\ell$ is smooth on the interatomic distance, the free-energy cost of long-wavelength fluctuations of the beating pattern can be described by a phenomenological expansion in terms of derivatives of the phason field, $\partial_i\tilde{u}_j$, constrained only by the symmetries of the continuum model. Since the superlattice vectors imposes a preferential orientation of the beating pattern (i.e., only translations and not rotations are soft\cite{foot_soft}), introducing a symmetric strain tensor is not sensible for this problem. Instead, $\partial_i\tilde{u}_j$ can be arranged in irreducible representations of $D_6$ (see Fig.~\ref{fig:symmetry}),\begin{subequations}
\label{eq:irreps}
\begin{align}
& \partial_x\tilde{u}_x+\partial_y\tilde{u}_y  \sim A_1,\\
& \partial_x\tilde{u}_y-\partial_y\tilde{u}_x  \sim A_2,\\
& \left[\begin{array}{c}
\partial_x\tilde{u}_x-\partial_y\tilde{u}_y\\
-\partial_x\tilde{u}_y-\partial_y\tilde{u}_x
\end{array}\right] \sim E_2,
\end{align}
\end{subequations}
representing compressional, rotational (or tilting), and shear deformations of the soliton network, respectively. A generic harmonic expansion reads then\begin{widetext}
\begin{align}
\mathcal{F}\left[\mathbf{\tilde{u}}\left(\mathbf{r}\right)\right]= \frac{1}{2}\int d\mathbf{r} & \left[\tilde{\lambda}\left(\boldsymbol{\nabla}\cdot\mathbf{\tilde{u}}\right)^2+\frac{\tilde{\mu}}{2}\left(\partial_i\tilde{u}_j+\partial_j\tilde{u}_i\right)^2+\gamma\left(\boldsymbol{\nabla}\times\mathbf{\tilde{u}}\right)^2\right],
 \label{eq:F}
\end{align}
\end{widetext}
where the cost of compressional and shear deformations is expressed in terms of new Lam\'e coefficients and $\gamma$ is the tilt modulus accounting for the cost of rotations with respect to the moir\'e superlattice. The elastic constants can be estimated from the approximated profile in Eq.~\eqref{eq:domain_wall} as (see Appendix~\ref{sec:A})\begin{subequations}
\label{eq:elastic_constants}
\begin{align}
& \tilde{\lambda}\left(\theta\right)\approx\frac{\left(1+\nu\right)\ell\, V\sin\left(\frac{\theta}{2}\right)}{a\left(\nu-1\right)},\\
& \tilde{\mu}\left(\theta\right)\approx\frac{\left(3-\nu\right)\ell\, V\sin\left(\frac{\theta}{2}\right)}{a\left(1-\nu\right)},\\
& \gamma\left(\theta\right)\approx\frac{4\ell\, V\sin\left(\frac{\theta}{2}\right)}{a\left(1-\nu\right)},
\end{align}\end{subequations}
where $\nu\approx 0.2$ is graphene's Poisson ratio.

The soliton network appears to have, within the limitations of the present calculation, a negative Poisson's ratio ($\tilde{\lambda}<0$).\cite{note_bulk} This is not so surprising for a two-dimensional incommensurate structure,\cite{Halperin_etal} and manifests the tendency of the resulting beating pattern to preserve the six-fold symmetry of the moir\'e superlattice. Here I should emphasize that a relative compression/expansion of the layers (like the uniaxial hetero-strain considered in Ref.~\onlinecite{fu_strain}) introduces a shear deformation of the beating pattern, and viceversa, relative shear between layers (which is energetically cheaper) introduces a longitudinal distortion. The latter can be interpreted as a modulation of the moir\'e period or a nonuniform twist angle, which is systematically observed in topographic images acquired by scanning tunneling microscopy.\cite{STM1,STM2,STM3} The lower energy for longitudinal deformations is ultimately ascribed to the fact that the soliton network connecting AA stacked regions must be understood as a system of strings under tension, whose energy scales linearly (instead of quadratically, like in a system of springs) with length. The tension of the domain walls manifest the metastability of bilayers with a nominal twist angle. These structures can be further stabilized by entropic terms in the free energy, not included in the purely mechanical model discussed here. Two-dimensional soliton configurations carry a lot of entropy,\cite{Villain,book} which is also associated with the fact that their energy scales linearly with length. Consequently, thermal fluctuations can contribute to renormalize the elastic constants. \cite{Pokrovsky_Tapalov,Halperin_etal} The existence of patches or domains with a nominal twist angle can also be interpreted as the fact that the transition from commensurate to incommensurate structures is first order and dominated by entropy.

Longitudinal and transverse phason modes are represented in Fig.~\ref{fig:fig2}~d~and~e, respectively. Their frequencies deduced from the harmonic expansion in Eq.~\eqref{eq:F} reads
\begin{subequations}
\label{eq:frequencies}
\begin{align}
& \omega_{\mathbf{q}}^{(L)}=\sqrt{\frac{\tilde{\lambda}+2\tilde{\mu}}{\tilde{\rho}}}\,\left|\mathbf{q}\right|\approx\sqrt{\frac{5-3\nu}{4\left(1-\nu\right)}}\,c\left|\mathbf{q}\right|,\\
& \omega_{\mathbf{q}}^{(T)}=\sqrt{\frac{\tilde{\mu}+\gamma}{\tilde{\rho}}}\,\left|\mathbf{q}\right|\approx\sqrt{\frac{7-\nu}{4\left(1-\nu\right)}}\,c\left|\mathbf{q}\right|,
\label{eq:transverse_mode}
\end{align}
\end{subequations}
where I have introduced in the mass density of the soliton network (see Appendix~\ref{sec:A}):
\begin{align}
\label{eq:solitom_mass_density}
\tilde{\rho}\left(\theta\right)\approx\rho\,\sqrt{\frac{8V}{\pi^2\mu}}\,\sin\left(\frac{\theta}{2}\right).
\end{align}
The inertia of the soliton system is reduced because the formation of sharper (on the scale of $L_M$) stacking textures implies that, effectively, a smaller fraction of the atoms within the moir\'e cell takes part of the sliding motion of one layer with respect to the other. This compensates the reductions of the elastic constants, so the dispersion of the phason modes do not change much with respect to the acoustic phonons of individual graphene layers, in agreement with numerical calculations.\cite{phonons_2013} This does not imply, as we have seen, that interlayer adhesion forces are negligible, or that the elasticity of individual layers can describe the energetics of phason modes.%, for which lattice relaxation is significant.

\section{Electron-phason coupling}

\label{sec:e-ph}

In order to evaluate the effect of phason fluctuations on the electronic spectrum and transport properties, I am going to consider the continuum model usually discussed in the literature:\cite{portu,macdonald}
\begin{align}
\label{eq:continuum_model}
\hat{\mathcal{H}}^{\left(\tau\right)}=\left(\begin{array}{cc}
\hat{\mathcal{H}}_D^{\left(\tau,t\right)} & \hat{T}^{\left(\tau\right)}\left(\mathbf{r}\right) \\
\left[\hat{T}^{\left(\tau\right)}\left(\mathbf{r}\right)\right]^{\dagger} & \hat{\mathcal{H}}_D^{\left(\tau,b\right)}
\end{array}\right).
\end{align}
The Hamiltonian is written in block form, each of them acting on a spinor wave function in a given valley sector (labelled by $\tau=\pm1$) from top and bottom layers (upper and lower blocks, respectively) describing electronic states around points $K_{\tau,\mu}$ in Fig.~\ref{fig:symmetry}. The block-diagonal terms are Dirac Hamiltonians of the form
\begin{align}
\hat{\mathcal{H}}_D^{\left(\tau,\mu\right)}=\hbar v_F\, \mathbf{\Sigma}^{\left(\tau,\mu\right)}\cdot\left(\hat{\mathbf{k}}-\mathbf{K}_{\tau,\mu}\right),
\end{align}
where $\hat{\mathbf{k}}=-i\boldsymbol{\partial}$ is the crystalline-momentum operator in real-space representation, $\hbar v_F=3t a/2$, with $t\approx 2.8$ eV being the intralayer hopping parameter, and $\mathbf{\Sigma}^{\left(\tau,\mu\right)}$ is a vector of Pauli matrices defined in the spinor (sublattice) space. Since crystalline momentum is expressed in a common frame of reference (defined by $\mathcal{C}_{2x}$ and $\mathcal{C}_{2y}$ axes in Fig.~\ref{fig:symmetry}) the Pauli matrices have to be properly rotated,
\begin{align}
\mathbf{\Sigma}^{\left(\tau,\mu\right)}=\left(\tau\, e^{\frac{i\mu\theta\hat{\ell}_z}{2}} \hat{\sigma}_x\, e^{-\frac{i\mu\theta\hat{\ell}_z}{2}}, e^{\frac{i\mu\theta\hat{\ell}_z}{2}} \hat{\sigma}_y\, e^{-\frac{i\mu\theta\hat{\ell}_z}{2}}\right),
\end{align}
where $\hat{\ell}_z=\tau\hat{\sigma}_z/2$ is the generator of spinor rotations (see Appendix~\ref{sec:B}) and $\mu=\pm 1$ for top/bottom layer blocks. Interlayer tunneling processes are described by the off-diagonal blocks, given by
\begin{align}
\label{eq:tunneling}
\hat{T}^{\left(\tau\right)}\left(\mathbf{r}\right)=t_{\perp}\left[ \hat{T}_0^{\left(\tau\right)} + \hat{T}_1^{\left(\tau\right)} e^{-i\tau\mathbf{G}_1\cdot\left(\mathbf{r}-\mathbf{\tilde{u}}\right)} + \hat{T}_2^{\left(\tau\right)} e^{i\tau\mathbf{G}_2\cdot\left(\mathbf{r}-\mathbf{\tilde{u}}\right)} \right],
\end{align}
where $t_{\perp}\approx 110$ meV parametrizes the strength of the interlayer hopping; matrices $\hat{T}_{i}^{(\tau)}$ contain the pertinent phases acquired by the wave function when electrons hop between different sublattices:
\begin{subequations}
\label{eq:matrices}
\begin{align}
& \hat{T}_0^{\left(\tau\right)}=\left(\begin{array}{cc}
1 & 1 \\
1 & 1
\end{array}\right),\\
& \hat{T}_1^{\left(\tau\right)}=\left(\begin{array}{cc}
1 & e^{i\tau\frac{2\pi}{3}} \\
e^{-i\tau\frac{2\pi}{3}} & 1
\end{array}\right),\\
& \hat{T}_2^{\left(\tau\right)}=\left(\begin{array}{cc}
1 & e^{-i\tau\frac{2\pi}{3}} \\
e^{i\tau\frac{2\pi}{3}} & 1
\end{array}\right).
\end{align}
\end{subequations}

In addition to time-reversal and superlattice-translation symmetries, along with valley conservation imposed by construction, the continuum model is also invariant under $D_6$ point-group operations\cite{symmetry1,symmetry2} and infinitesimal \textit{rigid} translations of one layer with respect to the other: a uniform phason field $\mathbf{\tilde{u}}$ parametrizing the center of the beating pattern can be absorbed in a unitary rotation of the wave function (specifically, it can be absorbed as a phase in the Bloch states defined in different copies of the moir\'e Brillouin zone), so the mini-band spectrum remains invariant. Note also that the model is derived assuming no lattice relaxation (the main steps are reproduced in Appendix~\ref{sec:B} following Ref.~\onlinecite{Koshino2}). Its effect can be incorporated by means of $i$) strain fields within the Dirac-Hamiltonian blocks and $ii$) more harmonics in $\hat{T}(\mathbf{r})$ along with smaller amplitudes for interlayer hoppings between the same sublattice, reflecting the shrinking of AA stacked areas. Although these terms have an important effect on the electronic spectrum,\cite{Koshino,Koshino3,Guinea_Walet} I am going to neglect them in my estimation of the electron-phason coupling; the observation is that, regardless of the changes in the Hamiltonian, the invariance under relative translations of the two layers imposes the parametric dependence on $\mathbf{\tilde{u}}$ already contained in Eq.~\eqref{eq:tunneling}. Thus, we can obtain the leading contribution in $t_{\perp}$ by expanding Eq.~\eqref{eq:tunneling} (and its hermitic conjugate) up to linear order in the smooth (on the scale of the moir\'e period) nonuniform phason field, just like in the case of a charge-density wave,\cite{cdw}
\begin{widetext}
\begin{align}
%\nonumber
\delta\hat{T}^{\left(\tau\right)}\left(\mathbf{r}\right)\approx &\, i\,\tau\, t_{\perp}\mathbf{\tilde{u}}\left(\mathbf{r}\right)\cdot \left[\mathbf{G}_1\,\hat{T}_1^{\left(\tau\right)} e^{-i\tau\mathbf{G}_1\cdot\mathbf{r}}
%\right. \\
%& \left.
 - \mathbf{G}_2\,\hat{T}_2^{\left(\tau\right)} e^{i\tau\mathbf{G}_2\cdot\mathbf{r}} \right].
 \label{eq:e-ph_1stq}
\end{align}
Introducing Fourier series for the phason field, decomposing the Fourier components in longitudinal and transverse components, and promoting the latter to boson operators in conventional fashion, provided that $\boldsymbol{\pi}=\tilde{\rho}\, \dot{\boldsymbol{\tilde{u}}}$ is the canonical momentum density conjugate to soliton displacements, I arrive at the following general expression for the electron-phason coupling in second quantization, \begin{align}
%\nonumber
\label{eq:e-ph_2ndq}
& \hat{H}_{\textrm{e-ph}}=-\sum_{\tau,\mathbf{k}',\mathbf{k},\mathbf{q}} \tau\, t_{\perp} \left\{
\sqrt{\frac{\hbar}{2 A\tilde{\rho}\omega_{\mathbf{q}}^{\left(L\right)}}}\, \hat{\textrm{A}}_{\mathbf{q}}^{\left(L\right)} \left[\frac{\mathbf{q}\cdot\mathbf{G}_1}{\left|\mathbf{q}\right|}\,\left(\Psi^{\left(\tau,t\right)}_{\mathbf{k}'-\tau\mathbf{G}_1}\right)^{\dagger}\hat{T}_1^{\left(\tau\right)}\Psi_{\mathbf{k}}^{\left(\tau,b\right)}-\frac{\mathbf{q}\cdot\mathbf{G}_2}{\left|\mathbf{q}\right|}\,\left(\Psi^{\left(\tau,t\right)}_{\mathbf{k}'+\tau\mathbf{G}_2}\right)^{\dagger}\hat{T}_2^{\left(\tau\right)}\Psi_{\mathbf{k}}^{\left(\tau,b\right)}\right]
\right.
\\
& \left.
+ \sqrt{\frac{\hbar}{2 A\tilde{\rho}\omega_{\mathbf{q}}^{\left(T\right)}}}\,\hat{\textrm{A}}_{\mathbf{q}}^{\left(T\right)}\left[\frac{\left(\mathbf{q}\times\mathbf{G}_1\right)_z}{\left|\mathbf{q}\right|}\,\left(\Psi^{\left(\tau,t\right)}_{\mathbf{k}'-\tau\mathbf{G}_1}\right)^{\dagger}\hat{T}_1^{\left(\tau\right)}\Psi_{\mathbf{k}}^{\left(\tau,b\right)}-\frac{\left(\mathbf{q}\times\mathbf{G}_2\right)_z}{\left|\mathbf{q}\right|}\,\left(\Psi^{\left(\tau,t\right)}_{\mathbf{k}'+\tau\mathbf{G}_2}\right)^{\dagger}\hat{T}_2^{\left(\tau\right)}\Psi_{\mathbf{k}}^{\left(\tau,b\right)}\right]
\right\}\delta_{\mathbf{k}',\mathbf{k}+\mathbf{q}}+\textrm{h.c.},
\nonumber
\end{align}
where $A$ is the area of the system, $\hat{\textrm{A}}_{\mathbf{q}}^{(\nu)}=a_{\mathbf{q}}^{\nu}+(a_{-\mathbf{q}}^{\nu})^{\dagger}$, and $\Psi_{\mathbf{k}}^{(\tau,\mu)}=(c_{A,\mathbf{k}}^{\tau,\mu},c_{B,\mathbf{k}}^{\tau,\mu})^T$; operators $(c_{\alpha,\mathbf{k}}^{\tau,\mu})^{\dagger}$/$c_{\alpha,\mathbf{k}}^{\tau,\mu}$ create/annihilate Bloch states with crystalline momentum $\mathbf{k}$ around valley $\tau$ in sublattice $\alpha$ of layer $\mu$, while $(a_{\mathbf{q}}^{\nu})^{\dagger}$/$a_{\mathbf{q}}^{\nu}$ creates/annihilates longitudinal ($\nu=L$) and transverse ($\nu=T$) phasons with momentum $\mathbf{q}$.
\end{widetext}

A similar coupling was considered in Ref.~\onlinecite{umklapp} to estimate the contribution from phonon umklapp scattering in graphene on boron nitride. In fact, Eq.~\eqref{eq:e-ph_2ndq} captures both normal and umklapp scattering processes when the reconstruction of the electronic spectrum is taken into account and the resulting mini-bands are represented in a reduced-zone scheme, with crystal momentum $\mathbf{k}$ restricted to the first moir\'e Brillouin zone. For low carrier concentration, umklapp processes are mediated by phonons with momenta of the order of $\mathbf{G}_i$ and, therefore, their contribution to the resistivity scales as $\varrho\propto e^{-\hbar\omega_{\mathbf{G}_i}/k_BT}/T$. In what follows, I am going to restrict the analysis to quasi-elastic scattering processes within the Fermi circles around the mini-Dirac points of the spectrum.

\subsection{Low-energy Hamiltonian}

\begin{figure*}[t!]
\begin{center}
%\hspace{-0.4cm}
\includegraphics[width=\textwidth]{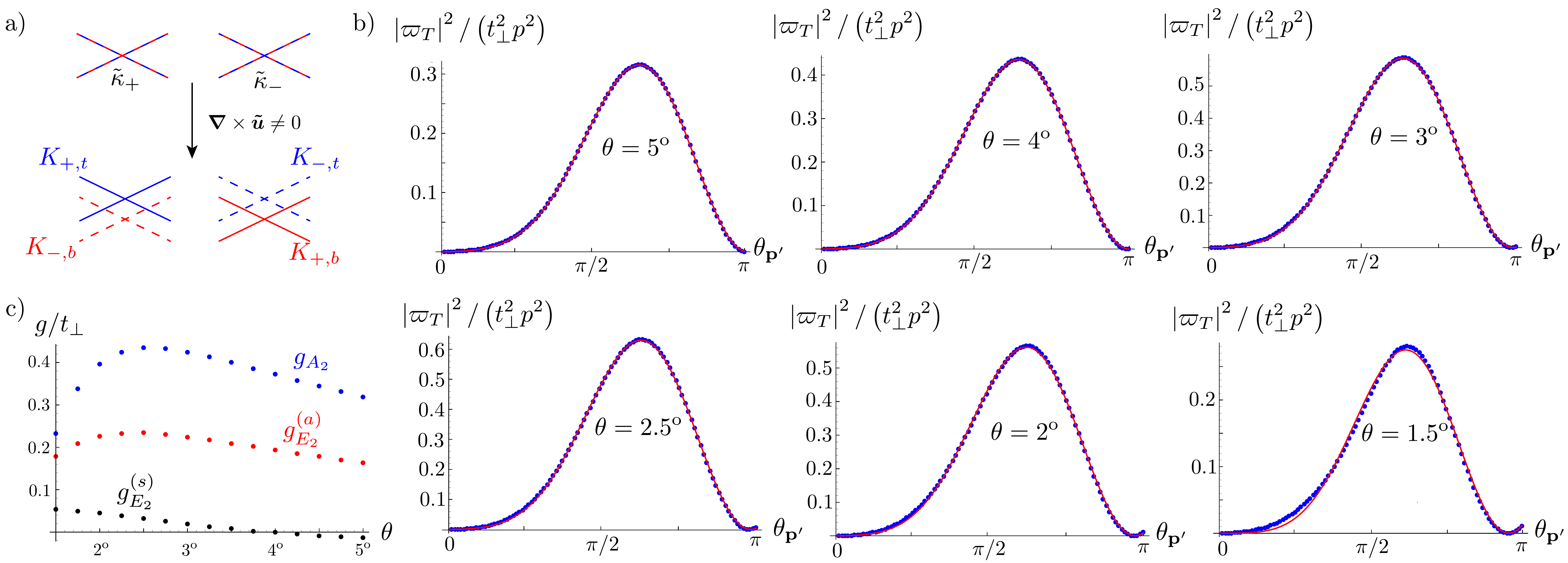}
\caption{a) Schematic representation of the splitting of the Dirac mini-bands under a local \textit{tilting} distortion of the moir\'e pattern (coupling parametrized by $g_{A_2}$); dashed lines correspond to valley $\tau=+1$, straight lines to valley $\tau=-1$, blue color to top layer $\mu=+1\,(t)$, and red color to bottom layer $\mu=-1\,(b)$. b) Electron-phason matrix elements in valley $\tau=+1$ within the conduction band $\zeta=+1$ around $\kappa_{+}$ point as a function of the scattering angle. The incident momentum $\mathbf{p}=p\,\mathbf{\hat{e}}_x$ corresponds to a filling of $n=10^{11}$ cm$^{-2}$ in the Dirac model, i.e., $pa=0.0056$. Blue dots correspond to the numerical evaluation in the continuum model for the electronic bands, red curves correspond to the phenomenological fitting in Eq.~\eqref{eq:fitting}. c) Electron-phason couplings within the low-energy Dirac bands as a function of the twist angle extracted from the fittings.}
%\vspace{-0.5cm} 
\label{fig:matrix_element}
\end{center}
\end{figure*}

Before tacking the calculation of the resistivity, let me discuss first the impact of phason fluctuations on the low-energy sector of the electronic spectrum. In a commensurate approximant, valleys $K_{\tau,\mu}$ are folded onto corners $\tilde{\kappa}_{\eta}$ of the moir\'e Brillouin zone; Figure~\ref{fig:symmetry} illustrates the case for type-I (in the nomenclature of Ref.~\onlinecite{symmetry2}, \textit{sublattice-exchange odd} in Ref.~\onlinecite{Mele0}) commensurate structures, which are dense in the limit of small angles.\cite{portu2} A $\mathbf{k}\cdot\mathbf{p}$ expansion around these points reads%\cite{foot5}
\begin{align}
\label{eq:kp}
\hat{\mathcal{H}}^{\left(\tau,\mu\right)}=\hbar v_F^*\left(\tau\hat{\sigma}_x,\hat{\sigma}_y\right)\cdot\mathbf{p}+\hat{\mathcal{H}}_{\textrm{e-ph}}^{\left(\tau,\mu\right)},
\end{align}
where $\mathbf{p}$ is the crystalline momentum around $\tilde{\kappa}_{\eta}$. This Hamiltonian acts on a new spinor basis for each valley sector $\tau$, each entry corresponding to envelope wave functions mostly localized on a given sublattice of layer $\mu$; the hybridization with other sublattices/layers is measured by the parameter $\alpha\equiv t_{\perp}/(\hbar v_F|\tilde{\kappa}_{\eta}|)$. The new Fermi velocity can be estimated in perturbation theory\cite{macdonald} as $v_F^*/v_F=(1-3\alpha^2)/(1+6\alpha^2)$; $\alpha^2=1/3$ defines the first magic angle in this approximation.

The second term in Eq.~\eqref{eq:kp} represents the electron-phason coupling projected onto the lowest-energy bands. The most general phenomenological expansion allowed by symmetry reads
\begin{widetext}
\begin{align}
\nonumber
\label{eq:couplings}
\hat{\mathcal{H}}_{\textrm{e-ph}}^{\left(\tau,t/b\right)}=\, & g_{A_1}\boldsymbol{\nabla}\cdot\mathbf{\tilde{u}}\,\pm g_{A_2}\left(\boldsymbol{\nabla}\times\mathbf{\tilde{u}}\right)_z+g_{E_2}^{(s)}\left[ \left(\partial_x\tilde{u}_x-\partial_y\tilde{u}_y\right) \hat{\sigma}_x - \tau\left(\partial_x\tilde{u}_y+\partial_y\tilde{u}_x\right)\hat{\sigma}_y\right]
\\
& \pm g_{E_2}^{(a)}\left[\tau \left(\partial_x\tilde{u}_x-\partial_y\tilde{u}_y\right)\hat{\sigma}_y + \left(\partial_x\tilde{u}_y+\partial_y\tilde{u}_x\right)\hat{\sigma}_x\right],
\end{align}
\end{widetext}
where the upper/lower sign applies to top/bottom layer sectors. Note that the electron-phason coupling can only depend on derivatives of the phason field due to the invariance of the electronic spectrum with respect to infinitesimal translations of the moir\'e pattern. The combinations in Eq.~\eqref{eq:irreps} can be paired with electronic operators transforming under the same irreducible representation of $D_6$ to form invariants under the point group and time reversal operations; details can be found in Appendix~\ref{sec:B}. The coupling constants $g_i$ are phenomenological parameters with units of energy.

%\begin{figure}[t!]
%\begin{center}
%\hspace{-0.4cm}
%\includegraphics[width=0.8\columnwidth]{scheme.pdf}
%\caption{Schematic representation of the splitting of the Dirac mini-bands under a shear or local twist distortion of the moir\'e pattern; dashed lines correspond to valley $\tau=+1$, straight lines to valley $\tau=-1$, blue color to top layer $\mu=+1\,(t)$, and red color to bottom layer $\mu=-1\,(b)$.}
%\vspace{-0.5cm} 
%\label{fig:scheme}
%\end{center}
%\end{figure}

The first and third terms in Eq.~\eqref{eq:couplings} resemble the \textit{scalar} and \textit{vector} electron-phonon couplings in graphene and are expected to be subleading in layer hybridization, $g_{A_1},g_{E_2}^{(s)}\sim\mathcal{O}(\alpha^2)$; consequently, the electron-phason coupling is dominated by transverse modes. The other two couplings act with opposite signs on valleys $K_{\tau,\pm}$ coming from different layers. In particular, tilting the soliton network ($\boldsymbol{\nabla}\times\mathbf{\tilde{u}}\neq 0$) with respect to the preferential direction imposed by the moir\'e superlattice lifts the degeneracy of these points, as illustrated in Fig.~\ref{fig:matrix_element}~a; the coupling parametrized by $g_{E_s}^{(a)}$ includes the effect of distortions of the beating pattern that break the 3-fold rotational symmetry, displacing the positions of the Dirac crossings in $\mathbf{k}$-space. These couplings reproduce the effect of relative strain between layers in the band structure\cite{fu_strain} and, as noted before, could explain the reduced Landau level degeneracy reported in magnetotransport.

Figure~\ref{fig:matrix_element}~b shows the numerical evaluation (blue dots) of the matrix elements of the coupling with transverse phason modes in the second line of Eq.~\eqref{eq:e-ph_2ndq}:\begin{align}
\left\langle \zeta',\tau,\mathbf{k}' \right|\hat{H}_{\textrm{ep}}\left|\zeta,\tau,\mathbf{k}\right\rangle=\sqrt{\frac{\hbar}{2 A\tilde{\rho}\omega_{\mathbf{q}}^{\left(T\right)}}}\,\varpi_{T,\tau}^{\zeta,\zeta'}\left(\mathbf{q},\mathbf{k},\mathbf{k}'\right)\delta_{\mathbf{k}',\mathbf{k}+\mathbf{q}}.
\end{align}
Here $|\zeta,\tau,\mathbf{k}\rangle$ represents electronic states from valley $\tau$ in minin-band $\zeta$ with crystalline momentum $\mathbf{k}$ restricted to the first moir\'e Brillouin zone, which are obtained by diagonalizing the Hamiltonian in Eq.~\eqref{eq:continuum_model}; the resulting matrix is truncated to a finite number of momentum values in each layer coming from different copies of the moir\'e zone, ranging from $7$ for the largest twist angle ($\theta=5^{\textrm{o}}$) to $61$ for the smallest ($\theta=1.5^{\textrm{o}}$). The results are normalized by the incident momentum (measured with respect to the corresponding mini-Dirac point) along the $x$ axis corresponding to a filling of $n=10^{11}$ cm$^{-2}$ in the lowest-energy electron band and fitted (red curves) to the phenomenological expression derived from Eq.~\eqref{eq:couplings},\begin{widetext}\begin{align}
\label{eq:fitting}
\varpi_{T,\tau}^{\mu,\zeta}(\mathbf{q},\mathbf{p},\mathbf{p}')=-\mu\, g_{A_2}\left|\mathbf{q}\right|\cos\left(\frac{\theta_{\mathbf{p}'}-\mathbf{\theta}_{\mathbf{p}}}{2}\right)-\mu\tau\zeta\, g_{E_2}^{(a)}\left|\mathbf{q}\right|\cos\left(2\theta_{\mathbf{q}}+\frac{\theta_{\mathbf{p}'}+\mathbf{\theta}_{\mathbf{p}}}{2}\right)+\tau\zeta\, g_{E_2}^{(s)}\left|\mathbf{q}\right|\sin\left(2\theta_{\mathbf{q}}+\frac{\theta_{\mathbf{p}'}+\mathbf{\theta}_{\mathbf{p}}}{2}\right),
\end{align}\end{widetext}
from which I estimate the coupling constants shown in Fig.~\ref{fig:matrix_element}~c for different twist angles. The calculation is restricted to intra-band ($\zeta=\zeta'$), quasi-elastic processes, for which $|\mathbf{q}|=|\mathbf{p}'-\mathbf{p}|=2|\mathbf{p}|\sin(\frac{\theta_{\mathbf{p}'}-\theta_{\mathbf{p}}}{2})$. Note that in these last expressions momenta are measured with respect to $\tilde{\kappa}_{\eta}$, with $\eta=\mu\times\tau$ as prescribed by the folding scheme in Fig.~\eqref{fig:symmetry}, and $\theta_{\mathbf{k}}$ parametrizes the direction of momentum $\mathbf{k}$, namely, $\mathbf{k}=|\mathbf{k}|(\cos\theta_{\mathbf{k}},\sin\theta_{\mathbf{k}})$. In particular, the calculations of Fig.~\ref{fig:matrix_element} corresponds to $\tau=+1$ and $\mu=+1$; the values of the couplings do not change appreciably when the calculation is performed for different incident momenta or band/valley numbers within the low-energy Dirac cones. The accuracy of the fitting curves is very good down to angles of the order of $\theta=1.2^{\textrm{o}}$.

The layer-symmetric coupling $g_{E_2}^{(s)}$ is only appreciable for the smallest angles, as expected. The layer-asymmetric couplings are non-monotonic with twist angle. This behavior is reproduced by the $\mathbf{k}\cdot\mathbf{p}$ perturbative expansion presented in Appendix~\ref{sec:B}, which predicts a stronger coupling with \textit{tilting} deformations of the beating pattern; second-order perturbation theory gives
\begin{align}
\label{eq:gA2}
g_{A_2}\approx\frac{3\,\alpha\,t_{\perp}}{1+6\alpha^2}\left(\frac{v_F^*}{v_F}\right).
\end{align}
The coupling grows first as the layer hybridization of the electronic wave function $\alpha/(1+6\alpha^2)$ increases, but around $\theta\lesssim 3^{\textrm{o}}$ it starts to decrease following the same trend as the Fermi velocity.

%Note that I am not including here the effect of hetero-strain in the sample, which would contribute to this coupling through the deformation potential. 

\subsection{Phason-limited electronic transport}

Electron scattering by strong phason fluctuations is expected to contribute to the $T$-dependent resistivity of twisted bilayer graphene at small incommensurate angles. I am going to consider a semiclassical treatment in the framework of Boltzmann transport theory, which implicitly assumes that $k_F\ell\gg 1$, where $k_F$ is the momentum of carriers within the Fermi surface and $\ell$ is the phason-limited mean free path. The electron-phason coupling enters explicitly in the collision integral via a scattering probability rate, which can be computed from Fermi's golden rule as $\mathcal{W}_i^f=2\pi \hbar^{-1}|\langle f|\hat{H}_{\textrm{ep}}|i\rangle|^2\delta(E_f-E_i)$; %Here $\hat{H}_{\textrm{ep}}$ expresses the electron-phason coupling in second quantization, Eq.~\eqref{eq:e-ph_2ndq} in its most general form;
the initial state reads $|i\rangle=|\zeta,\tau,\mathbf{k}\rangle\otimes|n_{\mathbf{q}}^{(\nu)}\rangle$, %where $|\zeta,\tau,\mathbf{k}\rangle$ represents electronic states from valley $\tau$ in minin-band $\zeta$ with crystalline momentum $\mathbf{k}$ restricted to the first moir\'e Brillouin zone, and 
where $|n_{\mathbf{q}}^{(\nu)}\rangle$ represents a state with $n_{\mathbf{q}}^{(\nu)}$ phasons in branch $\nu$. Note that, although the electron-phason coupling projected onto the low-energy bands results from coherent superpositions of electronic states in the two layers, at this point we are neglecting inter-band coherences in the calculation of the resistivity, which could alter  the results as the system approaches the neutrality point.\cite{coherence1,coherence2} I am also going to assume that the phason ensemble thermalize much faster than electrons, so $n_{\mathbf{q}}^{(\nu)}$ reduces to a equilibrium Bose-Einstein distribution function. Phason emission/absorption processes scatterer the initial state into $|f\rangle=|\zeta',\tau,\mathbf{k}'\rangle\otimes|n_{\mathbf{q}}^{(\nu)}\pm 1\rangle$ at a rate\begin{align}
\nonumber
\mathcal{W}_{\zeta,\mathbf{k}}^{\zeta',\mathbf{k}'}= & \frac{2\pi}{\hbar}\sum_{\nu=L,T}\sum_{\mathbf{q}}\frac{\hbar\left|\varpi_{\nu,\tau}^{\zeta,\zeta'}\left(\mathbf{q},\mathbf{k},\mathbf{k}'\right)\right|^2}{2A\tilde{\rho}\omega_{\mathbf{q}}^{\left(\nu\right)}}\,\delta_{\mathbf{k}',\mathbf{k}+\mathbf{q}}\times\\
& \left[n_{\mathbf{q}}^{(\nu)}\,\delta\left(\varepsilon_{\zeta',\tau,\mathbf{k}'}-\varepsilon_{\zeta,\tau,\mathbf{k}}-\hbar\omega_{\mathbf{q}}^{\left(\nu\right)}\right)\right.
\\
&\left.
+\left(n_{\mathbf{q}}^{(\nu)}+1\right)\delta\left(\varepsilon_{\zeta',\tau,\mathbf{k}'}-\varepsilon_{\zeta,\tau,\mathbf{k}}+\hbar\omega_{\mathbf{q}}^{\left(\nu\right)}\right)
\right],\nonumber
\end{align}
%where $\varpi_{\nu,\tau}^{\zeta,\zeta'}(\mathbf{q},\mathbf{k},\mathbf{k}')$ represents the electronic part of the electron-phason coupling matrix element. The latter can be computed perturbatively for the lowest-energy bands as explained in Appendix~\ref{sec:B}. For intra-band, quasi-elastic processes, which are assumed to dominate the resistivity (at least at $T\ll T_F$, where $T_F=\hbar v_F ^* k_F$ is the Fermi temperature), it reads\begin{align}
%\label{eq:matrix_element}
%\varpi_{T,\tau}^{\mu=\pm1}(\mathbf{q},\mathbf{k},\mathbf{k}')=\mp g_{A_2}\left|\mathbf{q}\right|\cos\left(\frac{\theta_{\mathbf{k}'}-\mathbf{\theta}_{\mathbf{k}}}{2}\right),
%\end{align}
%where $g_{A_2}$ is the coupling with transverse phason modes given in Eq.~\eqref{eq:gA2} and $\theta_{\mathbf{k}}$ parametrizes the direction of momentum $\mathbf{k}$, namely, $\mathbf{k}=|\mathbf{k}|(\cos\theta_{\mathbf{k}},\sin\theta_{\mathbf{k}})$.

\begin{figure}[t!]
\begin{center}
%\hspace{-0.4cm}
\includegraphics[width=\columnwidth]{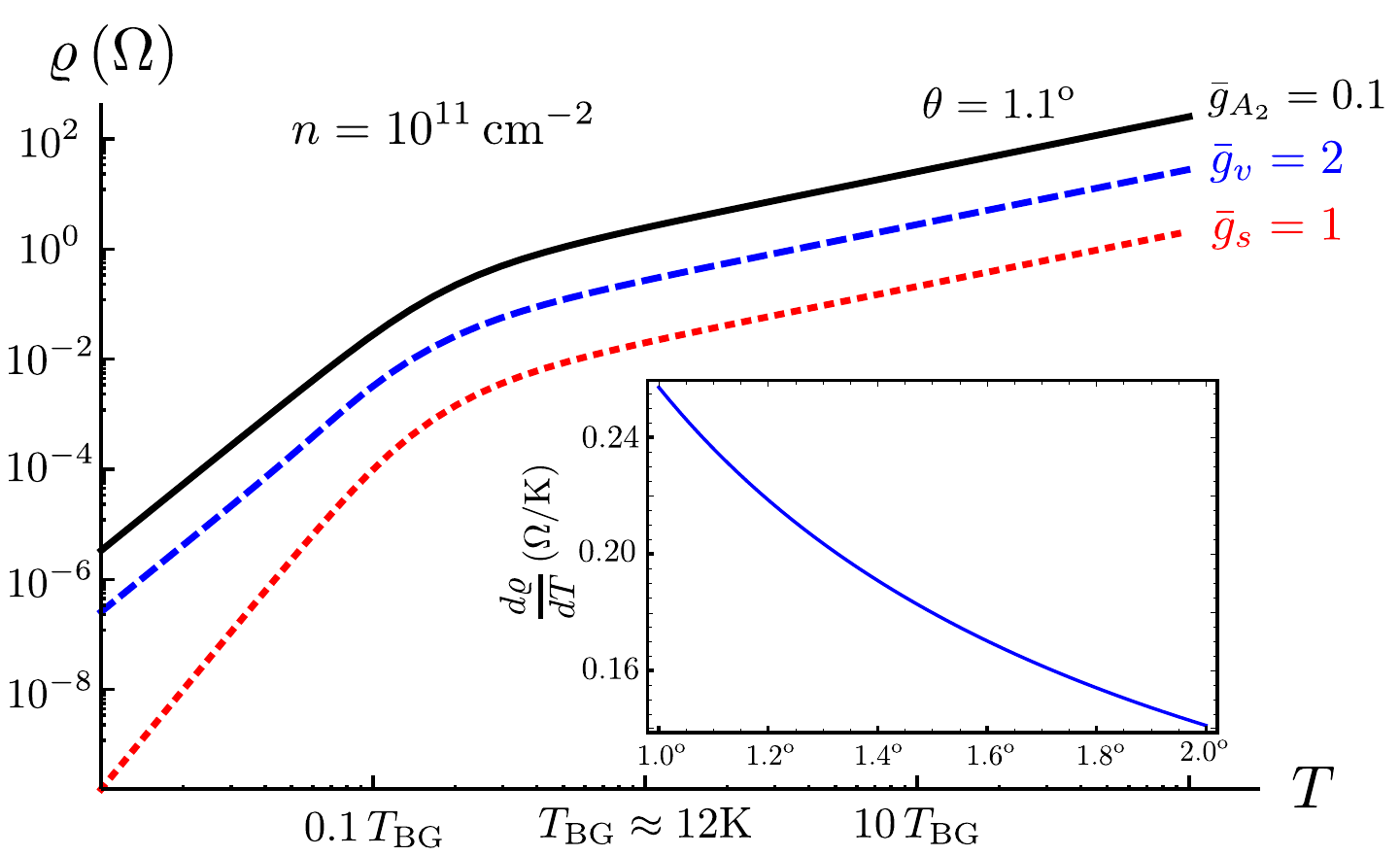}
\caption{$T$-dependent resistivity (in logarithmic scale) deduced from Eq.~\eqref{eq:resistivity} (black curve, corresponding to twist angle $\theta=1.1^{\textrm{o}}$). Dashed blue and dotted red curves show the contribution from the scalar ($\bar{g}_s$) and vectorial ($\bar{g}_v$) couplings with in-plane (layer symmetric) phonons.\cite{phonons_bilayer} The inset shows the dependence of $d\varrho/d T$ with twist angle in the high-$T$ regime.
}
%\vspace{-0.5cm} 
\label{fig:resistivity}
\end{center}
\end{figure}

In order to obtain analytical formulas, I am going to restrict the analysis to intra-band, quasi-elastic processes within the Dirac cones, dominated by the $g_{A_2}$ coupling with  transverse phason modes according to the estimates presented in the previous subsection. The resistivity can be obtained from a variational method\cite{Ziman} applied to the linearized Boltzmann equation describing the evolution of deviations of the electronic distribution function from equilibrium. The calculation is analogous to the case of the phonon-limited resistivity in graphene\cite{phonons_bilayer} and I am not going to reproduce the details here. The final result reads\begin{align}
\label{eq:resistivity}
\varrho=\frac{\hbar}{e^2}\frac{\left(2k_Fg_{A_2}\right)^2}{\tilde{\rho}\left(v_F^*\right)^2k_B T}\,I\left(\frac{T}{T_{\textrm{BG}}}\right),
\end{align}
where $I(x)$ is a dimensionless function defined by\begin{align}
I\left(x\right)=\int_0^1 dy\, y^4\sqrt{1-y^2}\,\frac{e^{y/x}}{\left(e^{y/x}-1\right)^2}.
\end{align}
The temperature scale $T_{\textrm{BG}}$, akin to the Bloch-Gr\"uneisen temperature in the problem of electron-phonon scattering, is related to the maximum momentum transfer ($2k_F$) between electronic states in a quasi-elastic scattering event, \begin{align}
k_BT_{\textrm{BG}}=\hbar \omega_{2k_F}^{\left(T\right)}.
\end{align}
At temperatures much lower than $T_{\textrm{BG}}$, the resistivity is dominated by small-angle scattering events and scales as $\varrho\sim(T/T_{\textrm{BG}})^4$. In the high-temperature regime, $T\gg T_{\textrm{BG}}$, the resistivity grows linearly with $T$ as\begin{align}
\label{eq:resisitvity_asymptotic}
\varrho \approx\frac{\pi\left(g_{A_2}\right)^2k_B T}{16e^2\hbar\left(v_F^*\right)^2\left(\tilde{\mu}+\gamma\right)}.
\end{align}
Given the dispersion relation in Eq.~\eqref{eq:transverse_mode} and the relation between the Fermi momentum and the carrier concentration in the low-energy Dirac model, $k_F=\sqrt{\pi n/2}$, the crossover takes place at temperatures of the order of $T_{\textrm{BG}}\approx 12\sqrt{n}$ K, with $n$ measured in units of $10^{11}$ cm$^{-2}$. Figure~\ref{fig:resistivity} (black curve) shows the dependence on temperature for fixed values of the carrier density ($n=10^{11}$ cm$^{-2}$) and the dimensionless electron-phason coupling $\bar{g}_{A_2}\equiv ag_{A_2} /\hbar v_F^*= 0.1$.

\section{Discussion}

\label{sec:discussion}

A linear-$T$ resistivity is expected in general if electron transport is limited by scattering off boson fluctuations above a certain temperature scale, the latter defined by phase-space constraints in these scattering events. In particular, for the coupling with either phasons or layer-symmetric phonons (corresponding to acoustic in-phase vibrations of both layers, for which the crossover temperature is of the same order as $T_{\textrm{BG}}$ defined above), the slope of the resistivity with temperature can be written as
\begin{align}
\frac{d\varrho_i}{dT}\approx R_0\left(\bar{g}_i\right)^2\frac{k_B}{16 a^2\mathcal{K}_i},
\end{align}
where $R_0=\pi\hbar/e^2$ is the quantum of resistance, $\mathcal{K}_i$ is the suitable elastic modulus ($\tilde{\mu}+\gamma$ for transverse phasons, $\lambda+2\mu$ and $\mu$ for longitudinal and transverse phonons, respectively), and the dimensionless parameter $\bar{g}_i$ measures the strength of the coupling within the mini-Dirac bands in units of the bandwidth parameter $\hbar v_F^*/a$. In the case of phasons, we have $\bar{g}_{A_2}\sim t_{\perp}/t$, where the exact fraction depends on the grade of layer hybridization in the wave function and, therefore, is expected to be sensitive to the effect of lattice relaxation on the electronic spectrum, which is not taken into account in the estimates of Sec.~\ref{sec:e-ph}. Regarding layer-symmetric phonons, since the vector potential couples directly to the electron velocity operator, interference of the electronic wave function leads to the same cancelation as in the Fermi velocity, giving $\bar{g}_{v}\approx\beta\equiv-\partial \ln t/\partial \ln a\sim 2-3$. In the case of the scalar coupling, the same cancelation takes place due to the reconstruction of the electronic spectrum, which enters through the electrostatic screening of the deformation potential;\cite{Ziman} in a Thomas-Fermi treatment,\cite{phonons_bilayer} the dimensionless coupling reduces to\begin{align}
\bar{g}_{s}\approx\frac{aD}{8 e^2 k_e}\sim 1,
\end{align}
where $D\approx 20-30$ eV and $k_e=1/4\pi\epsilon_0$ are the bare deformation potential\cite{Ando} and Coulomb constants, respectively.

The conclusions of this analysis are the following: $i$) The effective couplings (normalized by the bandwidth) within the reconstructed Dirac cones should not depend strongly on the twist angle with the exception of, maybe, the coupling with phason modes due to its sensitivity to layer hybridization. $ii$) Despite the weaker coupling with phasons, these modes can dominate the resistivity at small twist angles due to the reduced stiffness of the stacking domain-wall system. The inset of Fig.~\ref{fig:resistivity} shows the dependence of the slope of the resistivity as a function of the twist angle prescribed by the scaling of the elastic constants in Eqs.~\eqref{eq:elastic_constants} (i.e., neglecting changes in the effective couplings). The resistivity increases as the twist angle decreases due to the reduction of the rigidity of the soliton network, but the slope is at least two orders of magnitude smaller than those reported in the experiments.\cite{phonons_transport1,phonons_transport2} This points to a different mechanism, possibly related to strong electron correlations around the magic angle. Recently, Gonz\'alez and Stauber have argued that perfect nesting for fillings close to the emergent van Hove singularity at the band edge gives rise to a marginal Fermi liquid scaling of the quasiparticle lifetime.\cite{Gonzalez_Stauber1} Partially related to this observation, a recent model for the linear-$T$ resistivity in the normal state of cuprate superconductors invoking umklapp scattering\cite{cuprates} starts from the assumption of a Fermi surface reconstruction to maximize commensurate nesting; this could be induced by a spin-wave instability that, also in the present case,\cite{phonons_Fu,Gonzalez_Stauber2} can compete with the superconducting order. %The broken 3-fold symmetry of the charge distribution in real space reported in tunneling spectroscopy\cite{STM1,STM2,STM3} points to the same direction.
Other indications are the fact that the low-temperature crossover is systematically smaller than $T_{\textrm{BG}}$, as pointed out in Ref.~\onlinecite{phonons_transport2}, and the absence of saturation of the resistivity around the Fermi temperature (only when higher-energy bands start to be populated the resistivity drops), which may indicate that electronic quasiparticles are not well defined.

%The fact that there are no appreciable changes in the slope of the resistivity when $T$ crosses $T_F$ and the resistivity only saturates when higher-energy bands start to be populated indicates that most likely electronic quasi-particles are not well defined. No saturation. Gonzalez and Stauber . FS reconstruction, STM.

The conclusions presented here are based on Boltzmann transport theory, which does not include mesoscopic effects ascribed to the intrinsically disordered nature of the devices. In particular, in the comparison with the contribution from conventional phonons, I have neglected the effect of disorder on the beating pattern, which could pin the soliton network and open a gap in the phason spectrum. The reduction of the effective rigidity of the soliton network implies that phason fluctuations are enhanced and, consequently, the moir\'e pattern is also more sensitive to perturbations induced by the substrate. The simplest perturbation to the harmonic theory in Eq.~\eqref{eq:F} is a \textit{weak} (in the sense of Larkin\cite{Larkin}) disorder potential of the form
\begin{align}
\label{eq:model_disorder}
V_{dis}=-\int d\mathbf{r}\,\mathbf{f}\left(\mathbf{r}\right)\cdot\mathbf{\tilde{u}}\left(\mathbf{r}\right).
\end{align}
Here $\mathbf{f}\left(\mathbf{r}\right)$ represent forces acting independently on the stacking solitons. This potential breaks the translational invariance of the incommensurate lattice and pin the soliton system. Positional order in the moir\'e superlattice is lost at distances of the order of \begin{align}
\label{eq:pinning}
L_c\approx\frac{6\left(\tilde{\lambda}+2\tilde{\mu}\right)}{\pi\sigma_{f}},
\end{align}
where $\sigma_f^2\equiv\overline{f^2}$ represents the dispersion in the distribution of forces. In the case of encapsulated samples, for example, we can estimate this parameter as $\sigma_f\sim V_{\textrm{BN}}/L_{\textrm{BN}}$, where $V_{\textrm{BN}}\approx0.1$ eV/nm$^2$ characterizes the adhesion energy between graphene and boron nitride,\cite{Katsnelson_numerics} and $L_{\textrm{BN}}\approx 14$ nm is the characteristic size of the moir\'e due to the lattice mismatch, about $1.8\%$. According to Eqs.~\eqref{eq:elastic_constants}~and~\eqref{eq:pinning}, around the magic angle $\theta\sim 1^{\textrm{o}}$ positional order in the corresponding moir\'e superlattice is lost at distances $L_c\sim 26$ nm, about twice the size of the moir\'e period; $L_c$ collapses to $L_M$ at angles $\theta\sim0.7^{\textrm{o}}$.

Equation~\eqref{eq:pinning} must be interpreted as a collective pinning length\cite{Larkin2,Lee} below which the soliton network responds elastically or, more accurately, the harmonic expansion in Eq.~\eqref{eq:F} holds. What happens beyond that length it is difficult to say due to the evident shortcomings of the model in Eq.~\eqref{eq:model_disorder}, which, for example, neglects the quasi-periodicity of the beating pattern and, possibly, of the interaction with boron nitride if the latter is aligned with the sample. The difficulty arises from the fact that realistic disorder potentials vary on length scales much shorter than the moir\'e period. These considerations will be taken into account in a future study. Nevertheless, it is worth emphasizing that, although these are only tentative estimates, it is precisely the low stiffness of the soliton network defining the beating pattern at small twist angle what makes the system so sensitive to structural disorder, explaining the widespread presence of \textit{twist angle disorder} in the samples.

Finally, another possible mechanism for the softening of the soliton network is the presence of strong nematic fluctuations of electronic origin, as suggested by tunneling spectroscopy.\cite{STM1,STM2,STM3} This observation is particularly relevant for the nematicity around the charge neutrality point, which seems to be more pronounced at stacking domain walls.\cite{STM2,STM3} The phenomenological Hamiltonian presented in Eq.~\eqref{eq:couplings} provides a good description of the coupling with phason fluctuations close to neutrality. The symmetry analysis presented in Appendix B can be used to construct phenomenological theories describing the coupling of nematic order parameters with lattice degrees of freedom. The interplay between electrons, phasons, and nematic fluctuations is also of potential relevance to understand the $T$-dependent resistivity observed in twisted bilayer graphene. On the experimental front, it would be interesting to have a systematic study of the anisotropy in transport as a function of carrier density and in the absence and presence of applied tensions (although this could be affected by other mesoscopic effects). It would be also very useful to compare the resistivity of devices under different hydrostatic pressure (which have been shown to display a similar phenomenology than \textit{magic angle} bilayers\cite{columbia}) in order to determine the dominant role of phasons in transport, since the application of pressure can pin the soliton network and suppress this contribution to the resistivity.

%At scales longer than $L_c$, the soliton network will still behave collectively as an elastic manifold, but trapped in a random potential with many metastable states and, therefore, should manifest some kind of glassy behavior. In addition to that, the interaction with the substrate can give rise to inhomogeneous strains \textit{embedded} within the soliton network. This form of disorder does not break the translational invariance of the soliton system but favors the formation of dislocations in the atomic rearrangements, which can also unbind thermally.\cite{Halperin_etal}

In conclusion, long-wavelength fluctuations of a moir\'e beating pattern in the limit of small twist angles are dominated by phason modes. Their contribution to resistivity grows linearly with $T$, with increasing slope as the twist angle decreases due to the reduction of the stiffness of the soliton network. This contribution alone, however, seems to be insufficient to explain the fast growth of the resistivity when the \textit{magic angle} is approached, pointing to a different mechanism that might involve the presence of strong nematic fluctuations or a Fermi surface reconstruction linked to the correlated phenomena at lower temperatures.

\acknowledgments

I would like to thank Francisco Guinea, Abhay Pasupathy, Matthew Yankowitz, and Cory Dean for valuable discussions, and Ricardo Zarzuela for his careful reading of an early version of this manuscript. This work has been supported by the NSF MRSEC program DMR-1420634. 

\vspace{0.2cm}

\textit{Note}: While finishing this manuscript, I came across a recent preprint,\cite{preprint} which presents a full calculation of the spectrum of oscillation with the account of lattice relaxation, following the same recipe as in Appendix~\ref{sec:A}. This work highlights the role of stacking domain walls, reaching the same conclusions as in Sec.~\ref{sec:mechanics}, in particular, the scaling of the elastic constants of the moir\'e superlattice in Eqs.~\eqref{eq:elastic_constants}.

\appendix

\section{Mechanical model}

\label{sec:A}

Lattice relaxation is described by an elastic free energy of the form
\begin{align}
\label{eq:model0}
F=F_{\textrm{el}}+F_{\textrm{ad}}.
\end{align}
The first terms accounts for in-plane elastic distortions of graphene layers,
\begin{align}
\label{eq:elastic}
F_{\textrm{el}}=\sum_{\mu=t,b}\int d\mathbf{r}\,\left[\frac{\lambda}{2}\left(\boldsymbol{\nabla}\mathbf{u}^{(\mu)}\right)^2+\frac{\mu}{4}\left(\partial_iu_j^{(\mu)}+\partial_ju_i^{(\mu)}\right)^2\right],
\end{align}
where $\mathbf{u}^{(\mu)}$ describes displacements of unit cells in layer $\mu$ with respect to their equilibrium positions in the absence of interlayer forces. It is convenient to introduce the relative displacement, $\mathbf{u}=\mathbf{u}^{(t)}-\mathbf{u}^{(b)}$, and total displacement, $\mathbf{v}=\mathbf{u}^{(t)}+\mathbf{u}^{(b)}$. In the simplest approximation, the adhesion energy, second term in Eq.~\eqref{eq:model0}, is a functional of the former field only,
\begin{align}
F_{\textrm{ad}}=\int d\mathbf{r}\,\mathcal{V}_{\textrm{ad}}\left[\mathbf{r},\mathbf{u}\left(\mathbf{r}\right)\right],
\end{align}
where $\mathcal{V}_{\textrm{ad}}$ is the adhesion potential introduced in the main text. Equation~\eqref{eq:FK} comes from minimizing the free energy of variations with respect to $\mathbf{u}$. Dynamical equations are derived from the total Lagrangian $L=K-F$, where the kinetic energy reads\begin{align}
\label{eq:kinetic_energy}
K=\frac{\rho}{4}\int d\mathbf{r}\,\left[\dot{\mathbf{v}}^2+\dot{\mathbf{u}}^2\right].
\end{align}
Here $\rho=7.6\times 10^{-7}$ kg/m$^2$ is the mass density of individual graphene layers.

The first field $\mathbf{v}$ represents in-phase displacements of both layers and can be identified with the original acoustic phonons.\cite{phonons_bilayer} From this point on, I am going to focus on the dynamics of relative displacements, $\mathbf{u}$. Let me consider deviations from a metastable configuration, $\delta\mathbf{u}(t,\mathbf{r})=\mathbf{u}(t,\mathbf{r})-\mathbf{u}^{(0)}(\mathbf{r})$, where $\mathbf{u}^{(0)}(\mathbf{r})$ is a solution of Eq.~\eqref{eq:FK}. Plugging this ansatz into the previous equations gives
\begin{align}
\label{eq:model}
F_{\textrm{el}}\left[\mathbf{u}^{(0)}\right]+F_{\textrm{el}}\left[\delta\mathbf{u}\right]+F_{\textrm{mix}}\left[\delta\mathbf{u},\mathbf{u}^{(0)}\right]+F_{\textrm{ad}}\left[\delta\mathbf{u},\mathbf{u}^{(0)}\right],
\end{align}
where the first and second terms are just Eq.~\eqref{eq:elastic} evaluated with the metastable solution and the corresponding deviation, while the third term mixes both,\begin{align}
& F_{\textrm{mix}}\left[\delta\mathbf{u},\mathbf{u}^{(0)}\right]= \int d\mathbf{r}\,\left[\frac{\lambda}{2}\,\partial_iu_i^{(0)}\partial_j\delta u_j 
\right.\\
& \left.
+ \frac{\mu}{4} \left(\partial_i u_j^{(0)}
 +\partial_j u_i^{(0)}\right)\left(\partial_i\,\delta u_j+\partial_j\delta u_i\right)\right].
 \nonumber
\end{align}
The last term comes from the adhesion energy; expanding up to quadratic order gives\begin{align}
\nonumber
F_{\textrm{ad}}\left[\delta\mathbf{u},\mathbf{u}^{(0)}\right]\approx & \int d\mathbf{r}\,\left\{\mathcal{V}_{\textrm{ad}}\left[\mathbf{r},\mathbf{u}^{(0)}\right]+\delta u_i\frac{\partial\mathcal{V}_{\textrm{ad}}}{\partial u_i}|_{\mathbf{u}^{(0)}}
\right.\\
& \left. +\, \frac{1}{2}\,\delta u_i\delta u_j\,\frac{\partial^2\mathcal{V}_{\textrm{ad}}}{\partial u_i\partial u_j}|_{\mathbf{u}^{(0)}}
\right\}.
\end{align}
Hereafter repeated latin indices are summed up. Integration by parts (dropping boundary terms) leads to\begin{align}
F\approx & \, F_0+U\left[\delta\mathbf{u}\right]+\int d\mathbf{r}\,\delta u_i\left[\frac{\partial \mathcal{V}_{\textrm{ad}}}{\partial u_i}|_{\mathbf{u}^{(0)}}
\right.\\
& \left.
-\frac{\lambda}{2}\partial_i\partial_j u_j^{(0)}-\frac{\mu}{2}\partial_j\left(\partial_i u_j^{(0)}+\partial_ju_i^{(0)}\right)\right].
\nonumber
\end{align}
The last term is $0$ just from Eq.~\eqref{eq:FK}. The first term represents the free energy of the equilibrium solution,\begin{align}
\nonumber
F_0= \int d\mathbf{r} & \,\left\{\frac{\lambda}{4}\left(\boldsymbol{\nabla}\mathbf{u}^{(0)}\right)^2+\frac{\mu}{8}\left(\partial_iu_j^{(0)}
 +\,\partial_ju_i^{(0)}\right)^2
 \right.\\
& \left.
 +\, \mathcal{V}_{\textrm{ad}}\left[\mathbf{r},\mathbf{u}^{(0)}\left(\mathbf{r}\right)\right]\right\},
\end{align}
while $U[\delta\mathbf{u}]$ describes the spectrum of harmonic oscillations,\begin{align}
\nonumber
U\left[\delta\mathbf{u}\right]=& \,\frac{1}{2}  \int d\mathbf{r} \,\left\{\frac{\lambda}{2}\left(\boldsymbol{\nabla}\delta\mathbf{u}\right)^2+\frac{\mu}{4}\left(\partial_i\delta u_j
 +\,\partial_j\delta u_i\right)^2
  \right.\\
& \left.
 +\, \delta u_i\delta u_j\,\frac{\partial^2\mathcal{V}_{\textrm{ad}}}{\partial u_i\partial u_j}|_{\mathbf{u}^{(0)}}
\right\}.
\label{eq:U}
\end{align}

The Euler-Lagrange equations describing oscillations around the metastable state reads then\begin{align}
-\rho\,\delta \ddot{u}_i+\frac{\lambda+\mu}{2}\partial_i\partial_j \delta u_j+\frac{\mu}{2}\partial_j\partial_j \delta u_i=\delta u_j\,\frac{\partial^2\mathcal{V}_{\textrm{ad}}}{\partial u_i\partial u_j}|_{\mathbf{u}^{(0)}}.
\end{align}
By introducing Fourier series, \begin{align}
\delta u_i\left(t,\mathbf{r}\right)=\frac{1}{\sqrt{A}}\sum_{\mathbf{q}}\int \frac{d\omega}{2\pi}\,u_i\left(\omega,\mathbf{q}\right)\,e^{i\mathbf{q}\cdot\mathbf{r}-i\omega t},
\end{align}
the problem reduces to solve the following secular equation,\begin{align}
\nonumber
\rho\, \omega^2 u_i\left(\omega,\mathbf{q}\right)= & \left[\mu\left|\mathbf{q}\right|^2\delta_{ij}+\left(\lambda+\mu\right)q_i q_j\right] u_j\left(\omega,\mathbf{q}\right)
\\
& +\sum_{\mathbf{G}}K_{ij}\left(\mathbf{G}\right) u_j\left(\omega,\mathbf{q}-\mathbf{G}\right),
\end{align}
where I have introduced \begin{align}
K_{ij}\left(\mathbf{G}\right)=\frac{4}{\sqrt{3}L_M^2}\int_{\textrm{moir\'e}} d\mathbf{r}\,e^{-i\mathbf{G}\cdot\mathbf{r}}\,\frac{\partial^2\mathcal{V}_{\textrm{ad}}}{\partial u_i\partial u_j}|_{\mathbf{u}^{(0)}}.
\end{align}

\subsection{Floating layers}

\begin{figure}[t!]
\begin{center}
%\hspace{-0.4cm}
\includegraphics[width=\columnwidth]{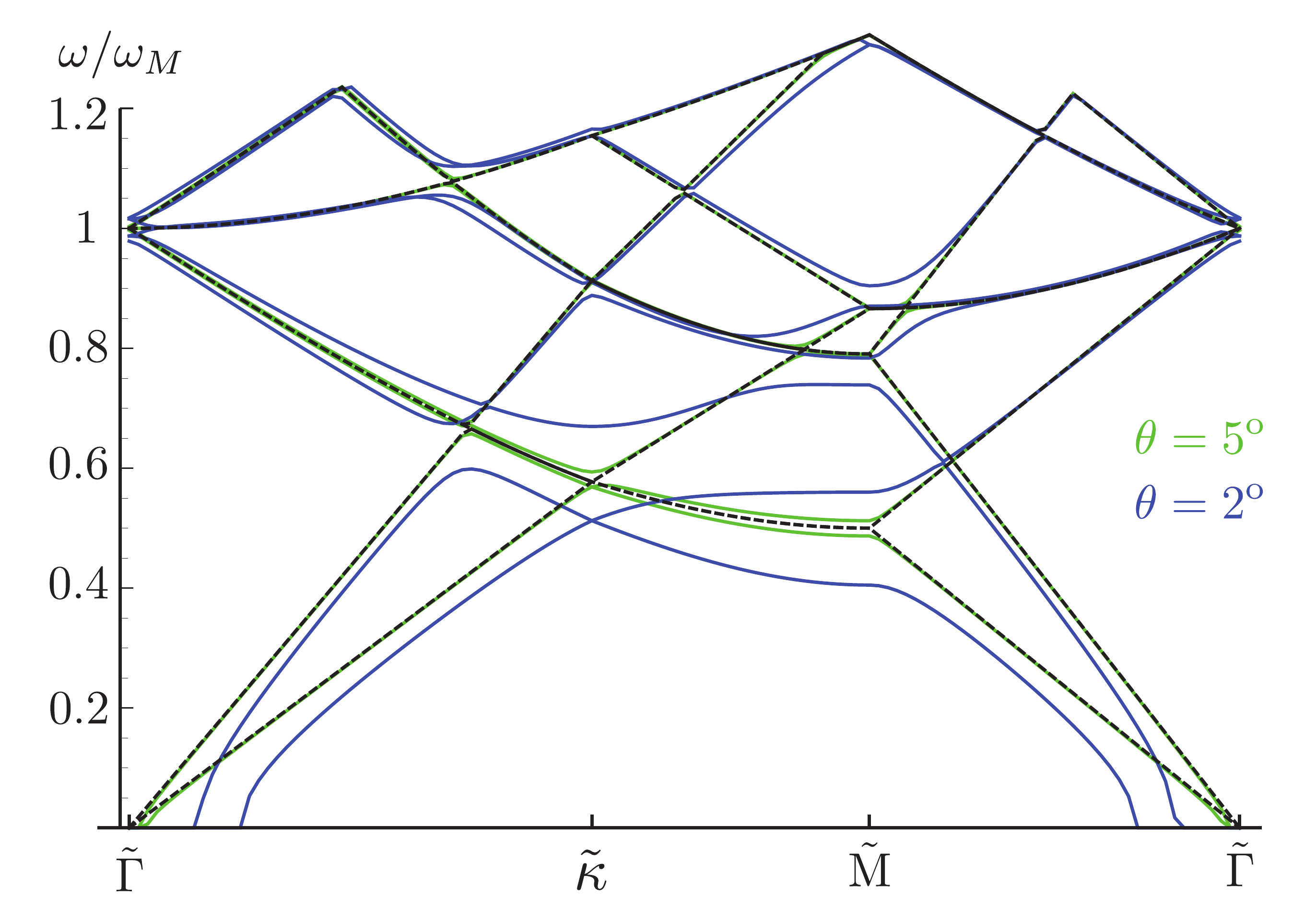}
\caption{Spectrum of oscillations of two floating layers. When $\ell$ and $L_M$ are comparable (for example, curves in green for $\theta=5^{\textrm{o}}$), the spectrum resembles the acoustic phonon branches of graphene folded onto the moir\'e Brillouin zone (in dashed black for reference). For smaller angles, the strong softening of the phonon modes around the zone center mark the instability of the system towards the formation of sharper stacking textures.}
%\vspace{-0.5cm} 
\label{fig:spectra}
\end{center}
\end{figure}

For large twist angles such that $L_M$ and $\ell$ (defined in Eq.~\ref{eq:length}) are comparable, we can neglect lattice relaxation, $\mathbf{u}^{(0)}\approx 0$, and consider oscillations around two floating layers. The average free energy per moir\'e cell is just $\bar{F}_0=3V/2$. The adhesion potential reduces to\begin{align}
\mathcal{V}_{\textrm{ad}}\left[\mathbf{r},0\right] = V\sum_{i=1}^3\left[\frac{1}{2}+\cos\left(\mathbf{G}_i\cdot \mathbf{r}\right)\right].
\label{eq:potential2}
\end{align}
In this approximation, all the harmonics $K_{ij}(\mathbf{G})$ are $0$ except for the ones in the first star ($\alpha=1,2,3$)\begin{align}
\label{eq:matrix_elements}
K_{ij}\left(\pm\mathbf{G}_{\alpha}\right) & =-\frac{V}{2} \left(\mathbf{b}_{\alpha}\right)_i\left(\mathbf{b}_{\alpha}\right)_j .
\end{align}
The spectra in Fig.~\ref{fig:spectra} are the result of truncating the secular equation to include the first 36 harmonics, giving a $37\times37$ matrix to diagonalize. The lowest-frequency branches are strongly softened (blue curves) when the matrix elements in Eq.~\eqref{eq:matrix_elements} start to be comparable with $\omega_M=4\pi c/(\sqrt{3}L_M)$.

\subsection{Soliton network}

Let me assume now that the lattice relaxes and the solution is roughly described by the superposition of the three density waves introduced in the main text,\begin{align}
\mathbf{u}^{(0)}\left(\mathbf{r}\right)=\sum_{\alpha=1,2,3}\mathbf{u}_{\alpha}\left(\varrho\right),
\end{align}
where $\mathbf{u}_{\alpha}\left(\varrho\right)\approx u\left(\varrho\right)\,\mathbf{\hat{u}}_{\alpha}$ is a soliton train along $\boldsymbol{\hat{\varrho}}_{\alpha}=\mathbf{\hat{z}}\times\mathbf{\hat{u}}_{\alpha}$, $u(\varrho)$ is given by Eq.~\eqref{eq:domain_wall}, and the unit vectors $\mathbf{\hat{u}}_{\alpha}$ lie along the three armchair directions: \begin{subequations}\begin{align}
& \mathbf{\hat{u}}_1=\left(-\frac{\sqrt{3}}{2},-\frac{1}{2}\right),\\
& \mathbf{\hat{u}}_2=\left(\frac{\sqrt{3}}{2},-\frac{1}{2}\right),\\
& \mathbf{\hat{u}}_3=\left(0,1\right).
\end{align}\end{subequations}
To the leading order in $\ell/L_M$, we can neglect soliton interactions at the crossings and approximate\begin{align}
\nonumber
\bar{F}_0 & \approx\frac{1}{\sqrt{3}L_M}\sum_{\alpha}\int d\varrho\left\{\frac{\mu}{2}\left(\partial_{\varrho} u_{\alpha}\right)^2+2\,V\left[0,\mathbf{u}_{\alpha}\left(\varrho\right)\right]\right\}\\
& \approx \frac{2\sqrt{3}\,\sigma}{L_M}.
\end{align}
For a sine-Gordon soliton, the tension reduces to \begin{align}
\sigma & = \int d\varrho\left\{\frac{\mu}{4}\left(\partial_{\varrho} u_{\alpha}\right)^2+V\left[0,\mathbf{u}_{\alpha}\left(\varrho\right)\right]\right\}\\
\nonumber
& =V\int d\varrho\,\text{sech}^2\left(\frac{\varrho-x_{\alpha}}{\ell}\right)=2\ell V=\frac{a}{\pi}\sqrt{2\mu V}.
\end{align}
%By comparing this estimate with the free-energy density of the floating state, we can see that the soliton network is preferred for twist angles such that $L_M>8\,\ell/\sqrt{3}$, corresponding to $\theta<0.95^{\textrm{o}}$ for the numerical values of the parameters considered in the main text. Note that already for bigger angles the softening of acoustic oscillations around the floating state is very pronounced. 

As argued in the main text, the acoustic branches in this limit can be identified with traveling-wave modes of the soliton network parametrized by collective coordinates $\mathbf{\tilde{u}}$. Specifically, if $\mathbf{u}^{(0)}\left(\mathbf{r}-\mathbf{\tilde{u}}\right)$ is the soliton-network solution centered at $\mathbf{\tilde{u}}$, then a smoothly distorted profile can be approximated by the functional \begin{align}
\mathbf{u}\left[\mathbf{\tilde{u}}\left(\mathbf{r}\right)\right]=\mathbf{u}^{(0)}\left(\mathbf{r}-\mathbf{\tilde{u}}\left(\mathbf{r}\right)\right),
\end{align}
where the collective coordinate has been promoted to a field $\mathbf{\tilde{u}}(\mathbf{r})$. Spatial derivatives can be approximated as\begin{align}
\partial_i u_j\approx\partial_i u_j^{(0)}-\partial_k u_j^{(0)}\partial_i\tilde{u}_k,
\end{align}
and therefore we can identify $\partial_i\delta u_j\approx -\partial_ku_j^{(0)}\partial_i\tilde{u}_k$. Plugging this result in the first line of Eq.~\eqref{eq:U} gives\begin{align}
& \mathcal{F}\left[\mathbf{\tilde{u}}\left(\mathbf{r}\right)\right]\approx\frac{1}{2}  \int d\mathbf{r} \left\{\frac{\lambda}{2}\,\partial_j u_i^{(0)}\partial_lu_k^{(0)}\left(\partial_i\tilde{u}_j\partial_k\tilde{u}_l\right)
\right.\\
& \left.
 +\frac{\mu}{2}\left[\partial_k u_j^{(0)}\partial_l u_j^{(0)}\left(\partial_i\tilde{u}_k\partial_i\tilde{u}_l\right)+\partial_k u_j^{(0)}\partial_lu_i^{(0)}\left(\partial_i\tilde{u}_k\partial_j \tilde{u}_l\right)\right]\right\}.
\nonumber
\end{align} 

We can now estimate from this last result the coefficients in the free-energy expansion of Eq.~\eqref{eq:F}. Since the spatial dependence of the phason field $\mathbf{\tilde{u}}(\mathbf{r})$ must be smooth on the scale of the moir\'e superlattice, the derivatives of $\mathbf{u}^{(0)}$ inside the integral can be approximated by their average over a moir\'e unit cell. We end up with\begin{align}
\mathcal{F}\left[\mathbf{\tilde{u}}\left(\mathbf{r}\right)\right]\approx\frac{C_{ijkl}}{2}\int d\mathbf{r}\,\partial_i\tilde{u}_j\partial_k\tilde{u}_l,
\end{align}
where \begin{align}
\label{eq:C}
C_{ijkl}= &  \frac{\lambda}{\sqrt{3}L_M^2}\int d\mathbf{r}\,\partial_ju_i^{(0)}\partial_l u_k^{(0)}
\\
& +\frac{\mu}{\sqrt{3}L_M^2}\int d\mathbf{r}\left\{\partial_ju_k^{(0)}\partial_l u_i^{(0)}+\delta_{ik}\,\partial_j u_m^{(0)}\partial_l u_m^{(0)}\right\}.
\nonumber
\end{align}
Note that the symmetry-adapted expansion in the main text adopts the same form, where the tensor of elastic coefficients reads\begin{align}
\label{eq:C_symmetry}
C_{ijkl}=\left(\gamma+\tilde{\mu}\right)\,\delta_{ik}\delta_{jl}+\tilde{\lambda}\,\delta_{ij}\delta_{kl}+\left(\tilde{\mu}-\gamma\right)\delta_{jk}\delta_{il}.
\end{align}

In order to evaluate the elastic constants, we can proceed as before and plug the superposition of the three soliton waves into Eq.~\eqref{eq:C}. Ignoring again the subleading contribution from soliton crossings, I find\begin{align}
\nonumber
C_{ijkl}= & I \sum_{\alpha}\left[\mu\,\delta_{ik}  \left(\mathbf{\hat{u}}_{\alpha}\right)_m\left(\mathbf{\hat{u}}_{\alpha}\right)_m\left(\boldsymbol{\hat{\varrho}}_{\alpha}\right)_j\left(\boldsymbol{\hat{\varrho}}_{\alpha}\right)_l
\right.
\\
& + \left.
\left(\lambda+\mu\right) \left(\mathbf{\hat{u}}_{\alpha}\right)_i\left(\mathbf{\hat{u}}_{\alpha}\right)_k\left(\boldsymbol{\hat{\varrho}}_{\alpha}\right)_j\left(\boldsymbol{\hat{\varrho}}_{\alpha}\right)_l\right],
\end{align}
where the prefactor reads\begin{align}
I=\frac{1}{\sqrt{3}L_M}\int d\varrho\left(\partial_{\varrho} u\right)^2=\frac{4 V\ell}{\sqrt{3}\mu L_M}.
\end{align}
The tensors between parenthesis are just\begin{widetext}\begin{subequations}\begin{align}
& \delta_{ik}  \sum_{\alpha}\left(\mathbf{\hat{u}}_{\alpha}\right)_m\left(\mathbf{\hat{u}}_{\alpha}\right)_m\left(\boldsymbol{\hat{\varrho}}_{\alpha}\right)_j\left(\boldsymbol{\hat{\varrho}}_{\alpha}\right)_l=\frac{3}{2}\delta_{ik}\delta_{jl}\\
& \sum_{\alpha} \left(\mathbf{\hat{u}}_{\alpha}\right)_i\left(\mathbf{\hat{u}}_{\alpha}\right)_k\left(\boldsymbol{\hat{\varrho}}_{\alpha}\right)_j\left(\boldsymbol{\hat{\varrho}}_{\alpha}\right)_l=\frac{9}{8}\delta_{ik}\delta_{jl}-\frac{3}{8}\delta_{ij}\delta_{kl}-\frac{3}{8}\delta_{jk}\delta_{il}.
\end{align}
\end{subequations}\end{widetext}
Comparing these expressions with Eq.~\eqref{eq:C_symmetry}, I arrive at the final formulas in Eqs.~\eqref{eq:elastic_constants} of the main text.

The effective mass density of the soliton network can be derived in the same manner by approximating the time derivatives in Eq.~\eqref{eq:kinetic_energy} as $\dot{u}_i\approx-\partial_ju_i^{(0)}\dot{\tilde{u}}_j$. The mass tensor of a soliton reads then\begin{align}
M_{ij}=\frac{\rho}{2}\int d^2\mathbf{r}\,\partial_iu_k^{(0)}\partial_ju_k^{(0)},
\end{align}
where the integral is extended over a moir\'e unit cell. Proceeding just as before, the final result reads $M_{ij}=M\delta_{ij}$, where $M$ is the inertia of a stacking domain wall,\begin{align}
M=\frac{3 a^2\rho L_M}{2\pi^2\ell}.
\end{align}
Dividing this quantity by the area of the moir\'e unit cell $A_M=\sqrt{3}L_M^2/2$ gives the mass density $\tilde{\rho}$ in Eq.~\eqref{eq:solitom_mass_density}. The dispersion relations in Eqs.~\eqref{eq:frequencies} follow from the corresponding Euler-Lagrange equations, where the Fourier components of the phason field can be decomposed in longitudinal and transverse components as usual,\begin{align}
\mathbf{\tilde{u}}\left(\mathbf{q}\right)=\frac{i\,\mathbf{q}}{\left|\mathbf{q}\right|}\,\tilde{u}_L\left(\mathbf{q}\right) + \frac{i\,\mathbf{\hat{z}}\times\mathbf{q}}{\left|\mathbf{q}\right|}\,\tilde{u}_T\left(\mathbf{q}\right).
\end{align}
These can be promoted to phason creation/annihilation operators in conventional fashion via the identification (here $\nu=L,T$ labels the branch) \begin{align}
\tilde{u}_{\nu}\left(\mathbf{q}\right)\longrightarrow\sqrt{\frac{\hbar}{2\,\tilde{\rho}\,\omega_{\mathbf{q}}^{(\nu)}}}\left[a_{\mathbf{q}}^{(\nu)}+\left(a_{-\mathbf{q}}^{(\nu)}\right)^{\dagger}\right],
\end{align}
where phason operators satisfy the boson algebra $[a_{\mathbf{q}_1},(a_{\mathbf{q}_2})^{\dagger}]=\delta_{\mathbf{q}_1,\mathbf{q}_2}$, following from the conjugacy relations $\{\tilde{u}_i(\mathbf{r}),\pi_j (\mathbf{r}')\}=\delta_{ij}\,\delta(\mathbf{r}-\mathbf{r}')$, where $\boldsymbol{\pi}=\tilde{\rho}\,\dot{\boldsymbol{\tilde{u}}}$.

\section{Electronic model}

\label{sec:B}

Let me start by writing a generic tight-binding Hamiltonian for two graphene layers floating on top of each other, $\hat{H}_{\textrm{tb}}=\hat{H}_{b}+\hat{H}_{t}+\hat{T}+\hat{T}^{\dagger}$. The first two terms represent intralayer hopping processes; up to first nearest neighbors, we have\begin{align}
\nonumber
\hat{H}_{\mu}= & -t\sum_i\left\{\left|\mathbf{R}_{i,A}^{(\mu)},\right\rangle\left\langle \mathbf{R}_{i,B}^{(\mu)}\right| + \left|\mathbf{R}_{i,A}^{(\mu)}\right\rangle\left\langle \mathbf{R}_{i,B}^{(\mu)}-\mathbf{a}_{1}^{(\mu)}\right| 
\right. \\
& \left.
+ \left|\mathbf{R}_{i,A}^{(\mu)}\right\rangle\left\langle \mathbf{R}_{i,B}^{(\mu)}-\mathbf{a}_{2}^{(\mu)}\right|  \right\}+\textrm{h.c.}
\end{align}
The Hamiltonian is written in a monoelectronic basis of Wannier $\pi$-orbitals localized on sites $A/B$ of unit cell $i$ of layer $\mu$. We can introduce Bloch states as\begin{align}
\left|\mathbf{R}_{i,\alpha}^{(\mu)}\right\rangle=\frac{1}{\sqrt{N}}\sum_{\mathbf{k}_{\mu}\in\textrm{BZ}_{\mu}} e^{-i\mathbf{k}_{\mu}\cdot\mathbf{R}_{i,\alpha}^{(\mu)}}\left|\mathbf{k}_{\mu},\alpha\right\rangle,
\end{align}
with crystalline momenta $\mathbf{k}_{\mu}$ restricted to the first Brillouin zone of layer $\mu$ ($N$ is the number of unit cells on each layer that I assume the same). The previous Hamiltonian reduces to\begin{align}
\hat{H}_{\mu}=\sum_{\mathbf{k}_{\mu}\in\textrm{BZ}_{\mu}}\sum_{\alpha,\beta=A,B}\left[\hat{H}_{\mathbf{k}_{\mu}}\right]_{\alpha,\beta}\left|\mathbf{k}_{\mu},\alpha\right\rangle \left\langle \mathbf{k}_{\mu},\beta \right|,
\end{align}
where the matrix in sublattice space reads\begin{align}
\label{eq:Hk}
\hat{H}_{\mathbf{k}_{\mu}}=-t\sum_{i=1}^3\left(\begin{array}{cc}
0 &  e^{i\mathbf{k}_{\mu}\cdot\boldsymbol{\delta}_i^{(\mu)}}\\
 e^{-i\mathbf{k}_{\mu}\cdot\boldsymbol{\delta}_i^{(\mu)}} & 0
\end{array}\right).
\end{align}
The sum is extended over the three vectors connecting $A$ with nearest B sites in layer $\mu$; in my notation, $\boldsymbol{\delta}_{1,2}^{(\mu)}=\boldsymbol{\delta}_{3}^{(\mu)}-\mathbf{a}_{1,2}^{(\mu)}$, and $\boldsymbol{\delta}_{3}^{(\mu)}=\mathbf{R}_{i,B}^{(\mu)}-\mathbf{R}_{i,A}^{(\mu)}$ connects the two sites within the unit cell.

The terms $\hat{T}$ ($\hat{T}^{\dagger}$) describe interlayer tunneling processes,\begin{align}
\label{eq:T}
\hat{T}=\sum_{i,j}\sum_{\alpha,\beta} T_{\alpha\beta}^{ij}\,\left|\mathbf{R}_{i,\alpha}^{(t)}\right\rangle\left\langle\mathbf{R}_{j,\beta}^{(b)}\right|.
\end{align}
In a two-center, Slater-Koster-like approximation, $T_{\alpha\beta}^{ij}$ depends only on the relative distance between Wannier centers, so it must admit a Fourier expansion of the form\begin{align}
T_{\alpha\beta}^{ij}=A_c\int \frac{d\mathbf{q}}{\left(2\pi\right)^2}\, e^{i\mathbf{q}\cdot\left(\mathbf{R}_{i,\alpha}^{(t)}-\mathbf{R}_{j,\beta}^{(b)}\right)}\,T\left(\mathbf{q}\right),
\end{align}
where $A_c$ is graphene's unit-cell area. Introducing Bloch states and plugging this last expression into Eq.~\eqref{eq:T}, we can rewrite the latter as\begin{widetext}\begin{align}
\label{eq:T_general}
\hat{T}=\sum_{\left\{ \mathbf{b}^{(b)} \right\}} \sum_{\left\{ \mathbf{b}^{(t)} \right\}}\sum_{\mathbf{k}_t,\mathbf{k}_b}\sum_{\alpha,\beta}e^{i\mathbf{b}^{(t)}\cdot\boldsymbol{\delta}_{\alpha}^{(t)}-i\mathbf{b}^{(b)}\cdot\boldsymbol{\delta}_{\beta}^{(b)}} T\left(\mathbf{k}_b+\mathbf{b}^{(b)}\right)\,\delta_{\mathbf{k}_b+\mathbf{b}^{(b)},\mathbf{k}_t+\mathbf{b}^{(t)}} \left|\mathbf{k}_{t},\alpha\right\rangle \left\langle \mathbf{k}_{b},\beta \right|,
\end{align}
where the first two sums are on reciprocal vectors of the top and bottom lattices. In deriving this expression, I have made use of the identities\begin{subequations}\begin{align}
& \sum_i e^{i\left(\mathbf{q}-\mathbf{k}_t\right)\cdot\mathbf{R}_{i,\alpha}^{(t)}}=N\,e^{i\mathbf{b}^{(t)}\cdot\boldsymbol{\delta}_{\alpha}^{(t)}}\,\delta_{\mathbf{q},\mathbf{k}_t+\mathbf{b}^{(t)}},\\
& \sum_j e^{-i\left(\mathbf{q}-\mathbf{k}_b\right)\cdot\mathbf{R}_{j,\beta}^{(b)}}=N\,e^{-i\mathbf{b}^{(b)}\cdot\boldsymbol{\delta}_{\beta}^{(b)}}\,\delta_{\mathbf{q},\mathbf{k}_b+\mathbf{b}^{(b)}},
\end{align}
\end{subequations}
where $\boldsymbol{\delta}_{\alpha}^{(\mu)}$ is the position of site $\alpha$ within a reference unit cell in layer $\mu$, such that $\boldsymbol{\delta}_{B}^{(\mu)}-\boldsymbol{\delta}_{A}^{(\mu)}=\boldsymbol{\delta}_{3}^{(\mu)}$.

Equation~\eqref{eq:T_general} describes scattering events satisfying the general umklapp condition\begin{align}
\label{eq:Umklapp}
\mathbf{k}_t+n_1'\mathbf{b}_1'+n_2'\mathbf{b}_ 2'=\mathbf{k}_b+n_1\mathbf{b}_1+n_2\mathbf{b}_ 2,
\end{align}
where $n_i$, $n_i'$ are integers. Here I have introduced the notation of the main text, namely, $\mathbf{b}_i$ and $\mathbf{b}_i'=R(\theta)\mathbf{b}_i$ are the primitive vectors of the reciprocal lattice of the bottom and top layers, respectively; similarly, I write $\boldsymbol{\delta}_{\alpha}^{(b)}\equiv \boldsymbol{\delta}_{\alpha}$, $\boldsymbol{\delta}_{\alpha}^{(t)}= R(\theta)\boldsymbol{\delta}_{\alpha}+\mathbf{u}$ where, in addition to the relative rotation along a common hexagon center, I consider a relative displacement $\mathbf{u}$ of the top with respect to the bottom layer. Commensurate approximants to the electronic structure simplifies the condition in Eq.~\eqref{eq:Umklapp} such that $n_i=n_i'$, i.e., $\mathbf{k}_t=\mathbf{k}_b+\mathbf{G}$, where $\mathbf{G}$ is a vector of the moir\'e superlattice; equation~\eqref{eq:T_general} simplifies then to\begin{align}
\label{eq:T_simplified}
\hat{T}\approx\sum_{\left\{ \mathbf{b} \right\}}\sum_{\mathbf{k}_b,\mathbf{k}_t}\sum_{\alpha,\beta}e^{i\mathbf{b}\cdot\left(\boldsymbol{\delta}_{\alpha}-\boldsymbol{\delta}_{\beta}\right)+i\mathbf{b}\cdot\left(R^T\mathbf{u}\right)}\, T\left(\mathbf{k}_b+\mathbf{b}\right) \delta_{\mathbf{k}_t,\mathbf{k}_b+(1-R)\mathbf{b}} \left|\mathbf{k}_{t},\alpha\right\rangle \left\langle \mathbf{k}_{b},\beta \right|.
\end{align}
The sum in $\left\{\mathbf{b}\right\}$ is also restricted in practice, provided that $T(\mathbf{q})$ is a rapidly decaying function of momentum; in particular, $T(\mathbf{q})$ is strongly suppressed for values $|\mathbf{q}|> 1/d$, where $d$ is the separation between layers.\cite{portu,macdonald}

For small twist angles, the electronic spectrum will be dominated by the low-energy Dirac bands of decoupled graphene layers lying around the two inequivalent corners (valleys) of the respective Brillouin zones labelled by $K_{\tau,\mu}$ in Fig.~\ref{fig:symmetry}; here $\tau=\pm 1$ labels the valleys, which is assumed to be a good quantum number. In a continuum description,\cite{portu,macdonald} band dispersion is simplified by expanding the phases in Eq.~\eqref{eq:Hk} for small momenta around $\mathbf{K}_{\tau,\mu}$,\begin{align}
\sum_{i=1}^3 e^{i\left(\mathbf{K}_{\tau,\mu}+\mathbf{p}_{\mu}\right)\cdot\boldsymbol{\delta}_i^{(\mu)}}\approx-\tau\, p_{\mu}^{(x)}+i p_{\mu}^{(y)},
\end{align}
where the $x$, $y$ components are adapted to the high-symmetry (zig-zag and armchair, respectively) axes of individual graphene layers. As mentioned in the main text, it is convenient to expresses crystalline momenta in a common frame of reference (defined in this case by the high-symmetry axes highlighted in Fig.~\ref{fig:symmetry}), so the spinor basis in the corresponding sublattice space of each layer has to be rotated, leading to the block-diagonal terms in Eq.~\eqref{eq:continuum_model}.

For the tunneling terms, we can, in the same spirit, neglect small deviations from $\mathbf{K}_{\tau,\mu}$ in the argument of $T(\mathbf{q})$. If we include only one harmonic, $t_{\perp}\equiv T\left(|\mathbf{K}_{\tau,\mu}|\right)$, the sum in $\{\mathbf{b}\}$ has to be restricted to $\mathbf{b}=\mathbf{0},-\tau\,\mathbf{b}_1,\tau\,\mathbf{b}_2$, corresponding to the three equivalent positions of a given valley within a single-layer Brillouin zone. Equation~\eqref{eq:T_simplified} simplifies to\begin{align}
\nonumber
\hat{T}\approx t_{\perp}\sum_{\mathbf{p},\tau}\sum_{\alpha,\beta}
& \left\{ \left[\hat{T}_{0}^{\left(\tau\right)}\right]_{\alpha,\beta} \left|\mathbf{k}_t=\mathbf{k}_b,\alpha\right\rangle \left\langle \mathbf{k}_b=\mathbf{K}_{\tau,b}+\mathbf{p},\beta \right| 
+ \left[\hat{T}_{1}^{\left(\tau\right)}\right]_{\alpha,\beta} e^{i\tau\mathbf{G}_1\cdot\mathbf{\tilde{u}}} \left|\mathbf{k}_{t}=\mathbf{k}_b-\tau\,\mathbf{G}_1,\alpha\right\rangle \left\langle \mathbf{k}_b=\mathbf{K}_{\tau,b}+\mathbf{p},\beta\right| 
\right.\\
& \left.
+ \left[\hat{T}_{2}^{\left(\tau\right)}\right]_{\alpha,\beta} e^{-i\tau\mathbf{G}_2\cdot\mathbf{\tilde{u}}} \left|\mathbf{k}_{t}=\mathbf{k}_b+\tau\,\mathbf{G}_2,\alpha\right\rangle \left\langle \mathbf{k}_b=\mathbf{K}_{\tau,b}+\mathbf{p},\beta\right| 
\right\},
\label{eq:T_continuum}
\end{align}
\end{widetext}
where I have used the relation in Eq.~\eqref{eq:distances} along with the identity in footnote~\onlinecite{foot1}, so that $\mathbf{b}\cdot(R^T\mathbf{u})=-\mathbf{G}\cdot\mathbf{\tilde{u}}$. Equation~\eqref{eq:T_continuum} transformed back to real space corresponds to Eq.~\eqref{eq:tunneling}, with matrices $\hat{T}_{i}$ given in Eqs.~\eqref{eq:matrices}.

\subsection{Symmetry-adapted electronic operators}

\begin{table*}[t!]
\centering
\begin{tabular}{|c||c||c|c|c|c|c|c|}
\hline
Electronic operators ($\mathcal{T}$) & Irrep & \, $\mathcal{E}$ \, & \, $\mathcal{C}_{2z}$\, & $2\,\mathcal{C}_{3z}$ & $2\,\mathcal{C}_{6z}$ & $3\,\mathcal{C}_{2x}$ & $3\,\mathcal{C}_{2y}$\\
\hline
\hline
$\hat{\mu}_z\hat{\tau}_z\hat{\sigma}_z\, (-)$ & $A_1$ & 1 & 1 & 1 & 1 & 1 & 1 \\
\hline
$\hat{\mu}_z\, (+)$, $\hat{\tau}_z\hat{\sigma}_z\, (-)$ & $A_2$ & 1 & 1 & 1 & 1 & -1 & -1 \\
\hline
$\hat{\tau}_z\, (-)$, $\hat{\mu}_z\hat{\sigma}_z\, (+)$ & $B_1$ & 1 & -1 & 1 &- 1 & 1 & -1 \\
\hline
$\hat{\sigma}_z\, (+)$, $\hat{\mu}_z\hat{\tau}_z\, (-)$ & $B_2$ & 1 & -1 & 1 & -1 & -1 & 1 \\
\hline
$\left(\begin{array}{c}
\hat{\tau}_z\hat{\sigma}_x \\
\hat{\sigma}_y \end{array}\right) (-)$,\, $\left(\begin{array}{c}
-\hat{\mu}_z\hat{\sigma}_y \\
\hat{\mu}_z\hat{\tau}_z\hat{\sigma}_x \end{array}\right) (-)$ & $E_1$ & 2 & -2 & -1 & 1 & 0 & 0 \\
\hline
$\left(\begin{array}{c}
\hat{\sigma}_x \\
\hat{\tau}_z\hat{\sigma}_y \end{array}\right) (+)$\,, $\left(\begin{array}{c}
- \hat{\mu}_z\hat{\tau}_z\hat{\sigma}_y \\
\hat{\mu}_z\hat{\sigma}_x \end{array}\right) (+)$ & $E_2$ & 2 & 2 & -1 & -1 & 0 & 0 \\
\hline
\end{tabular}
\caption{Classification of electronic valley- and layer-diagonal operators according to irreducible representations (irreps) of $D_6$ and parity ($\pm 1$, even/odd) under timer-reversal symmetry.}
\label{table}
\end{table*}

In order to construct phenomenological couplings within the low-energy sector of the spectrum, it is convenient to introduce a basis of electronic operators adapted to the irreducible representations (irreps) of the point-group symmetry $D_6$ of the continuum model. The internal Hilbert space of the continuum model is spanned by spin, sublattice, layer, and valley degrees of freedom. I am going to ignore the spin since relativistic corrections are weak. Note that neither sublattice nor layer are good quantum numbers, since inter-layer hopping terms mix them, but we can still refer to them to label the transformation properties of the Bloch wave function around $\tilde{\kappa}_{\eta}$ points. Valley $\tau=\pm 1$ and layer $\mu=\pm 1$ indices label four Dirac crossings reminiscent of the Dirac points of two decoupled graphene layers, while sublattice indices span the associated subspace of Dirac doublets at each crossing. %, two lying around $\tilde{\kappa}_{+}$ and the other two around $\tilde{\kappa}_{-}$ such that $\eta=\tau\times\mu$; the notation is summarized in Fig.~\ref{fig:symmetry}.
Electronic operators can be written in a basis of Pauli matrices acting on each subspace: $\hat{\sigma}_i$ for sublattice, $\hat{\tau}_i$ for valley, and $\hat{\mu}_i$ for layer. Since we are interested in the coupling to long-wavelength phason fluctuations on the scale of the moir\'e period, I am going to restrict the analysis to diagonal operators in valley and layer (or $\tilde{\kappa}_{\eta}$-point) numbers with the pertinent identifications $\hat{\tau}_z\rightarrow\tau$, $\hat{\mu}_z\rightarrow\mu$. I find convenient, however, to keep the matrix notation here, because the transformation rules of electronic operators are more easily identified from the Pauli-matrix algebra.

For example, elementary rotations along the 6-fold principal axis are implemented by unitary operators \begin{subequations}\begin{align}
& \mathcal{C}_{2z}: \hat{\tau}_x\hat{\sigma}_x,\\
& \mathcal{C}_{3z}: e^{\frac{i2\pi}{3}\hat{\ell}_z},
\end{align}
\end{subequations}
where $\hat{\ell}_z=\frac{1}{2}\hat{\tau}_z\hat{\sigma}_z$. Two-fold rotations along $x$ and $y$ axes are given by \begin{subequations}\begin{align}
& \mathcal{C}_{2x}: \hat{\mu}_x\hat{\sigma}_x,\\
& \mathcal{C}_{2y}: \hat{\mu}_x\hat{\tau}_x.
\end{align}
\end{subequations}
The rest of point-group operations follow from matrix multiplication. Finally, time-reversal symmetry is implemented by the anti-unitary operator\begin{align}
\mathcal{T}:\hat{\tau}_x\,\mathcal{K},
\end{align}
where $\mathcal{K}$ denotes complex conjugation.

All the possible combinations of valley- and layer-diagonal operators can be classified according to the irreps of $D_6$ and time-reversal symmetry. The result is summarized in Tab.~\ref{table}, along with the characters of $D_6$. It is worth emphasizing that there are two ways to lift the Dirac degeneracies without breaking time-reversal symmetry: by breaking both $\mathcal{C}_{2z}$ and $\mathcal{C}_{2y}$ symmetries ($\hat{\mu}_z\hat{\sigma}_z\sim B_1$), i.e., by removing the physical equivalence of both sublattices and layers (a staggered potential of opposite sign in each layer), or by breaking $\mathcal{C}_{2z}$ and $\mathcal{C}_{2x}$ symmetries ($\hat{\sigma}_z\sim B_2$), i.e., by removing only the physical equivalence between sublattices (the same staggered potential in both layers). Only the latter leads to bands with nonzero valley-Chern number. The competition between these two mass terms give rise to the phase diagram discussed in Ref.~\onlinecite{Zaletel}.

The phenomenological electron-phason Hamiltonian in Eq.~\eqref{eq:couplings} consists of all the possible invariants formed from combinations of the phason-field derivatives in Eqs.~\eqref{eq:irreps} with the corresponding electronic operators transforming under the same irrep. Here I should note that the phason field transforms as $\mathbf{\tilde{u}}=(\tilde{u}_x,\tilde{u}_y)\sim E_1$, while the relative displacement between layers transforms as $(-u_y,u_x)\sim E_1$ since, recall, $\mathcal{C}_{2x,y}$ rotations exchange the layers. The relation between these two transformation rules can be understood from the twist (second, dominant term) in Eq.~\eqref{eq:distances} or, with the account of lattice relaxation, the fact that solitons involve relative shear between layers. 

\subsection{Perturbative calculation of the electron-phason matrix element}

The Hamiltonian of the continuum model can be diagonalized in a basis of Bloch states by restricting the values of crystalline momentum to the first Brillouin zone of the moir\'e reciprocal lattice and introducing new (band) quantum numbers associated with different copies separated by momenta in $\{\mathbf{G}\}$. In this process, we can absorb a uniform phason field as a phase in the new electronic basis, so the spectrum remains invariant under translations of the moir\'e pattern. It is useful to consider the positions of the valleys folded into the first moir\'e Brillouin zone,
\begin{subequations}\begin{align}
& \mathbf{K}_{\pm,t}=\pm\frac{2\mathbf{G}_2+\mathbf{G}_1}{3}\left(\equiv\tilde{\kappa}_{\pm}\right),\\
& \mathbf{K}_{\pm,b}=\pm\frac{2\mathbf{G}_1+\mathbf{G}_2}{3}\left(\equiv\tilde{\kappa}_{\mp}\right).
\end{align}\end{subequations}
This folding scheme applies to type-I\cite{symmetry2} or sublattice-exchange odd\cite{Mele0} commensurate approximants, but once the first sum in Eq.~\eqref{eq:T_simplified} is restricted to one harmonic, this choice is inconsequential.\cite{portu2}

Next, we consider the dispersion of low-energy electronic states with momentum $\mathbf{p}$ around these points, in the previous notation, $\mathbf{k}_{t}=\mathbf{K}_{\tau,t}+\mathbf{p}$ (defining the top layer sector, $\mu=+1$), and $\mathbf{k}_{b}=\mathbf{K}_{\tau,b}+\mathbf{p}$ (corresponding to the bottom layer sector, $\mu=-1$). For each moir\'e Brillouin zone in a given layer I am going to truncate the number of copies in the opposite layer to three, so each (decoupled) sector labelled by valley and layer numbers is described by a $8\times 8$ matrix Hamiltonian of the form
\begin{align}
\label{eq:8x8}
\hat{\mathcal{H}}_{8\times8}^{\left(\tau,\mu\right)}=\hat{\mathcal{H}}_{\mathbf{p}}^{\left(\tau,\mu\right)}+\hat{\mathcal{U}}^{\left(\tau,\mu\right)}.
\end{align}
The first term reads just (in block form)\begin{widetext}\begin{align}
\hat{\mathcal{H}}_{\mathbf{p}}^{\left(\tau,+1\right)}=\left(\begin{array}{cccc}
\hbar v_F\, \mathbf{\Sigma}^{\left(\tau,+1\right)}\cdot\mathbf{p} & 0 & 0 &0\\
0 & \hbar v_F\, \mathbf{\Sigma}^{\left(\tau,-1\right)}\cdot\mathbf{p} & 0 & 0\\
0 &0 &  \hbar v_F\, \mathbf{\Sigma}^{\left(\tau,-1\right)}\cdot\mathbf{p} & 0\\
0 & 0 & 0 &  \hbar v_F\, \mathbf{\Sigma}^{\left(\tau,-1\right)}\cdot\mathbf{p}
\end{array}\right)
\end{align}
for the top-layer sector, and similarly for the bottom layer, \begin{align}
\hat{\mathcal{H}}_{\mathbf{p}}^{\left(\tau,-1\right)}=\left(\begin{array}{cccc}
\hbar v_F\, \mathbf{\Sigma}^{\left(\tau,+1\right)}\cdot\mathbf{p} & 0 & 0 &0\\
0 & \hbar v_F\, \mathbf{\Sigma}^{\left(\tau,+1\right)}\cdot\mathbf{p} & 0 & 0\\
0 &0 &  \hbar v_F\, \mathbf{\Sigma}^{\left(\tau,+1\right)}\cdot\mathbf{p} & 0\\
0 & 0 & 0 &  \hbar v_F\, \mathbf{\Sigma}^{\left(\tau,-1\right)}\cdot\mathbf{p}
\end{array}\right).
\end{align}
From this point on, I am going to neglect the rotation of the spinor basis, which restores the electron-hole symmetry of the spectrum (the error scales with $\theta^2$); while these matrices read the same now, note that they are expressed in a different basis. The second term in Eq.~\eqref{eq:8x8} reads\begin{align}
\hat{\mathcal{U}}^{\left(\tau,+1\right)}=\left(\begin{array}{cccc}
0 & t_{\perp}\hat{T}_0^{\left(\tau\right)} &  t_{\perp}\hat{T}_1^{\left(\tau\right)} & t_{\perp}\hat{T}_2^{\left(\tau\right)}\\
 t_{\perp}\hat{T}_0^{\left(\tau\right)} & \hbar v_F\, \mathbf{\Sigma}^{\left(\tau\right)}\cdot\left(\mathbf{K}_{\tau,t}-\mathbf{K}_{\tau,b}\right) & 0 & 0\\
 t_{\perp}\hat{T}_1^{\left(\tau\right)} &0 & \hbar v_F\, \mathbf{\Sigma}^{\left(\tau\right)}\cdot\left(\mathbf{K}_{\tau,t}-\mathbf{K}_{\tau,b}+\tau\mathbf{G}_1\right) & 0\\
 t_{\perp}\hat{T}_2^{\left(\tau\right)} & 0 & 0 &   \hbar v_F\, \mathbf{\Sigma}^{\left(\tau\right)}\cdot\left(\mathbf{K}_{\tau,t}-\mathbf{K}_{\tau,b}-\tau\mathbf{G}_2\right)
\end{array}\right)
\end{align}
for the top-layer sector, and\begin{align}
\hat{\mathcal{U}}^{\left(\tau,-1\right)}=\left(\begin{array}{cccc}
\hbar v_F\, \mathbf{\Sigma}^{\left(\tau\right)}\cdot\left(\mathbf{K}_{\tau,b}-\mathbf{K}_{\tau,t}\right) & 0 & 0 & t_{\perp}\hat{T}_0^{\left(\tau\right)} \\
0 & \hbar v_F\, \mathbf{\Sigma}^{\left(\tau\right)}\cdot\left(\mathbf{K}_{\tau,b}-\mathbf{K}_{\tau,t}-\tau\mathbf{G}_1\right) & 0 &  t_{\perp}\hat{T}_1^{\left(\tau\right)} \\
0 & 0 & \hbar v_F\, \mathbf{\Sigma}^{\left(\tau\right)}\cdot\left(\mathbf{K}_{\tau,b}-\mathbf{K}_{\tau,t}+\tau\mathbf{G}_2\right) & t_{\perp}\hat{T}_2^{\left(\tau\right)}\\
 t_{\perp}\hat{T}_0^{\left(\tau\right)} &  t_{\perp}\hat{T}_1^{\left(\tau\right)} & t_{\perp}\hat{T}_2^{\left(\tau\right)} & 0
\end{array}\right)
\end{align}
for the bottom-layer sector. The low-energy subspace is defined by the zero-energy eigenstates of $\hat{\mathcal{U}}^{\left(\tau,\mu\right)}$, specifically,\begin{subequations}\begin{align}
& \left|\psi_{A,\tau,+1}\right\rangle=\frac{1}{\sqrt{1+6\alpha^2}}\left(\begin{array}{c}
1\\
0\\
i\tau\alpha\\
-i\tau\alpha\\
i\tau\alpha\\
-i\tau\alpha \, e^{-i\tau\frac{2\pi}{3}}\\
i\tau\alpha\\
-i\tau\alpha \,e^{i\tau\frac{2\pi}{3}}
\end{array}\right), \,\,\,\,\,
 \left|\psi_{B,\tau,+1}\right\rangle=\frac{1}{\sqrt{1+6\alpha^2}}\left(\begin{array}{c}
0\\
1\\
i\tau\alpha\\
-i\tau\alpha\\
i\tau\alpha\, e^{i\tau\frac{2\pi}{3}}\\
-i\tau\alpha \\
i\tau\alpha \,e^{-i\tau\frac{2\pi}{3}}\\
-i\tau\alpha
\end{array}\right),
\\
& \left|\psi_{A,\tau,-1}\right\rangle=\frac{1}{\sqrt{1+6\alpha^2}}\left(\begin{array}{c}
-i\tau\alpha\\
i\tau\alpha\\
-i\tau\alpha\\
i\tau\alpha \, e^{-i\tau\frac{2\pi}{3}}\\
-i\tau\alpha\\
i\tau\alpha \,e^{i\tau\frac{2\pi}{3}}\\
1\\
0
\end{array}\right), \,\,\,\,\,
 \left|\psi_{B,\tau,-1}\right\rangle=\frac{1}{\sqrt{1+6\alpha^2}}\left(\begin{array}{c}
-i\tau\alpha\\
i\tau\alpha\\
-i\tau\alpha\, e^{i\tau\frac{2\pi}{3}}\\
i\tau\alpha \\
-i\tau\alpha \,e^{-i\tau\frac{2\pi}{3}}\\
i\tau\alpha\\
0\\
1
\end{array}\right).
\end{align}
\end{subequations}
These states form a basis for the $D_6$ irreps introduced in the previous subsection. The rest of eigenstates are separated by energies $E_i=\pm\hbar v_F|\tilde{\kappa}|$, $\pm\hbar v_F\sqrt{1+6\alpha^2}\,|\tilde{\kappa}|$ (the last ones are 2-fold degenerate in this approximation).

Next, I am going to produce a $\mathbf{k}\cdot\mathbf{p}$ expansion of the Hamiltonian by projecting-out the high-energy states in a L\"owdin perturbative scheme.\cite{Lowdin} Let me introduce first the following projection operators:\begin{subequations}\begin{align}
& \hat{P}_0^{\left(\tau,\mu\right)}\equiv\sum_{\alpha=A,B}\left|\psi_{\alpha,\tau,\mu}\right\rangle \left\langle\psi_{\alpha,\tau,\mu}\right|,\\
& \hat{1}-\hat{P}_0^{\left(\tau,\mu\right)}=\sum_i\hat{P}_i^{\left(\tau,\mu\right)},
\end{align}\end{subequations}
where $i$ labels the high-energy eigenstates. Up to second order in perturbation theory, where any term in the Hamiltonian of Eq.~\eqref{eq:8x8} (generically written as $\hat{V}^{(\tau,\mu)}$) is treated as a perturbation to $\hat{\mathcal{U}}^{\left(\tau,\mu\right)}$, we have \begin{align}
\label{eq:perturbative_scheme}
\hat{\mathcal{H}}^{\left(\tau,\mu\right)}\approx \hat{P}_0^{\left(\tau,\mu\right)}\hat{V}^{\left(\tau,\mu\right)}\hat{P}_0^{\left(\tau,\mu\right)}-\sum_i\hat{P}_0^{\left(\tau,\mu\right)}\hat{V}^{\left(\tau,\mu\right)}\frac{\hat{P}_i^{\left(\tau,\mu\right)}}{E_i}\hat{V}^{\left(\tau,\mu\right)}\hat{P}_0^{\left(\tau,\mu\right)}.
\end{align}
First-order perturbation theory in $\hat{\mathcal{H}}_{\mathbf{p}}^{\left(\tau,\mu\right)}$ leads to the first term in Eq.~\eqref{eq:kp} with a new Fermi velocity $v_F^*$ reduced by a factor $(1-3\alpha^2)/(1+6\alpha^2)$, the result first obtained by Bistritzer and MacDonald in Ref.~\onlinecite{macdonald}. For the electron-phason coupling, we should consider first its expression in the $8\times 8$ Hilbert space; from Eq.~\eqref{eq:e-ph_1stq}, we have\begin{align}
\hat{V}_{\textrm{e-ph}}^{\left(\tau,+1\right)}=\delta\hat{T}^{\left(\tau,+1\right)}+\textrm{h.c.}= \tau\, t_{\perp} \left(\begin{array}{cccc}
0 & 0 & i\,\mathbf{G}_1\cdot\mathbf{\tilde{u}}\,\, \hat{T}_1^{\left(\tau\right)} & -i\,\mathbf{G}_2\cdot\mathbf{\tilde{u}}\,\,\hat{T}_2^{\left(\tau\right)}\\
0 & 0 & 0 & 0\\
- i\,\mathbf{G}_1\cdot\mathbf{\tilde{u}}\,\,\hat{T}_1^{\left(\tau\right)} &0 & 0 & 0\\
 i\,\mathbf{G}_2\cdot\mathbf{\tilde{u}}\,\,\hat{T}_2^{\left(\tau\right)} & 0 & 0 & 0
\end{array}\right)
\end{align}
for the top-layer sector, and \begin{align}
\hat{V}_{\textrm{e-ph}}^{\left(\tau,-1\right)}=\delta\hat{T}^{\left(\tau,-1\right)}+\textrm{h.c.}= \tau\, t_{\perp} \left(\begin{array}{cccc}
0 & 0 & 0 & 0\\
0 & 0 & 0 & i\,\mathbf{G}_1\cdot\mathbf{\tilde{u}}\,\, \hat{T}_1^{\left(\tau\right)}\\
0 &0 & 0 & -i\,\mathbf{G}_2\cdot\mathbf{\tilde{u}}\,\, \hat{T}_2^{\left(\tau\right)}\\
0 &- i\,\mathbf{G}_1\cdot\mathbf{\tilde{u}}\,\,\hat{T}_1^{\left(\tau\right)} &  i\,\mathbf{G}_2\cdot\mathbf{\tilde{u}}\,\,\hat{T}_2^{\left(\tau\right)} & 0
\end{array}\right)
\end{align}
\end{widetext}
for the bottom layer. First-order perturbation theory gives $0$, as expected, since the invariance under translations of the soliton network implies that there must be momentum transfer between electronic states.\cite{foot_robustness} Treating both $\hat{V}_{\textrm{e-ph}}^{\left(\tau,\mu\right)}$ and $\hat{\mathcal{H}}_{\mathbf{p}}^{\left(\tau,\mu\right)}$ up to second order in perturbation theory gives\begin{align}
\label{eq:projected_e-ph}
\hat{\mathcal{H}}_{\textrm{e-ph}}^{\left(\tau,\mu\right)}\left(\mathbf{k},\mathbf{k}'\right) & \approx -\sum_i\hat{P}_0^{\left(\tau,\mu\right)}\hat{V}_{\textrm{e-ph}}^{\left(\tau,\mu\right)}\frac{\hat{P}_i^{\left(\tau,\mu\right)}}{E_i} \hat{\mathcal{H}}_{\mathbf{p}}^{\left(\tau,\mu\right)} \hat{P}_0^{\left(\tau,\mu\right)}+\textrm{h.c.}\nonumber\\
& \approx-\mu\tau\frac{6\,\alpha\, t_{\perp}\left(1-3\alpha^2\right)}{\left(1+6\alpha^2\right)^2}\left(\mathbf{p}\cdot\mathbf{\tilde{u}}\right)\hat{\sigma}_z,
\end{align}
where $\mathbf{p}$ must be interpreted now as the average momentum during a electron-phason scattering event, $\mathbf{p}\equiv(\mathbf{k}'+\mathbf{k})/2$, with $\mathbf{k}$, $\mathbf{k}'$ labelling the initial and final states, respectively.

\begin{figure}[t!]
\begin{center}
%\hspace{-0.4cm}
\includegraphics[width=0.9\columnwidth]{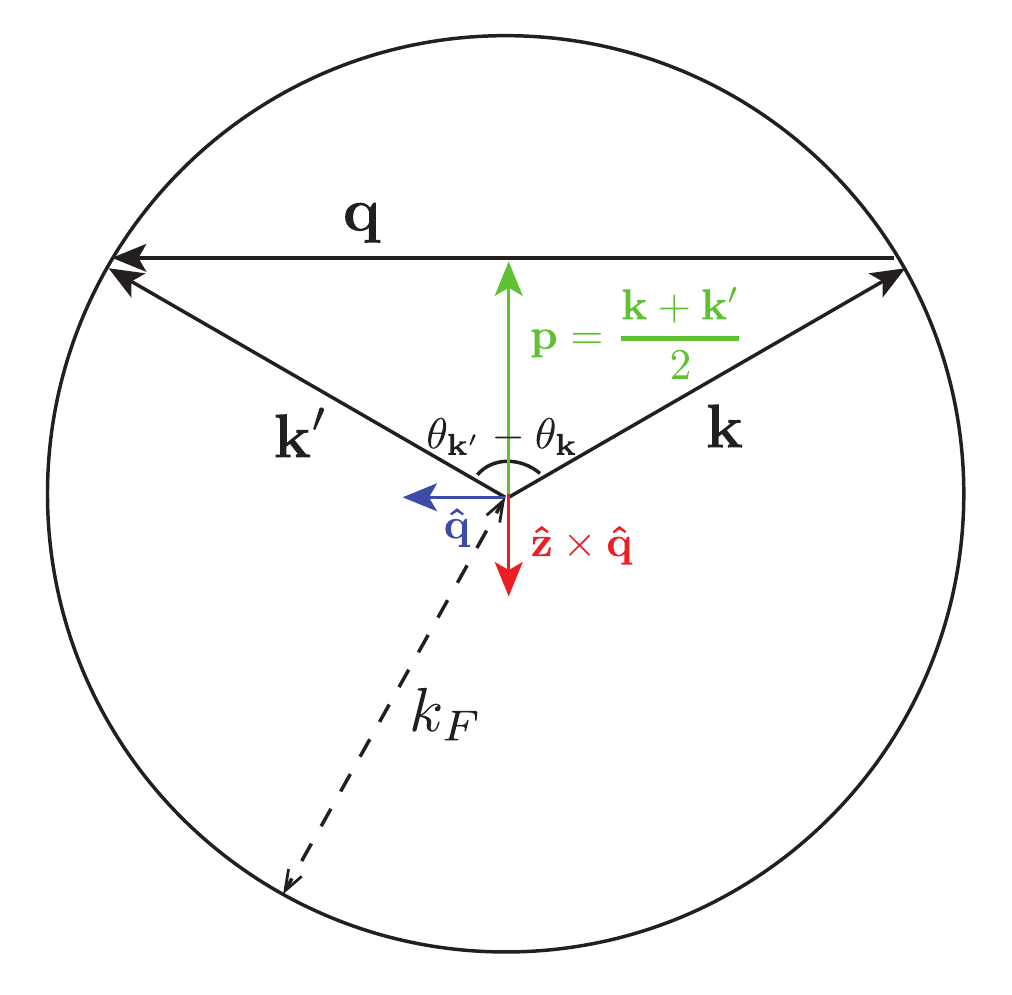}
\caption{Kinematics of quasi-elastic scattering events within a Fermi circle centered at one of the corners of the moir\'e Brillouin zone.}
%\vspace{-0.5cm} 
\label{fig:kinematics}
\end{center}
\end{figure}

Equation~\eqref{eq:projected_e-ph} may look odd at first glance, for it is difficult to recognize the symmetry-allowed couplings in the phenomenological expansion of Eq.~\eqref{eq:couplings} from this expression. Note, however, that Eq.~\eqref{eq:projected_e-ph} is compatible with time-reversal and $D_6$ point-group symmetries and, in fact, the combination of valley, sublattice and layer indices is such that when projected over a band state, i.e., and eigenstate of the first term in Eq.~\eqref{eq:kp}, \begin{align}
\left|\zeta,\tau,\mu,\mathbf{k}\right\rangle=\frac{e^{-\frac{i\tau\theta_{\mathbf{k}}}{2}}}{\sqrt{2}} \left|\psi_{A,\tau,\mu}\right\rangle +\tau\zeta \frac{e^{\frac{i\tau\theta_{\mathbf{k}}}{2}}}{\sqrt{2}} \left|\psi_{B,\tau,\mu}\right\rangle,
\end{align}
where $\zeta=\pm 1$ labels the electron/hole bands, Eq.~\eqref{eq:projected_e-ph} produces the expected matrix element of $g_{A_2}$ coupling. Note first that for intra-band processes we have\begin{align}
& \left\langle\zeta,\tau,\mu,\mathbf{k}'\right| \hat{\mathcal{H}}_{\textrm{e-ph}}^{\left(\tau,\mu\right)}\left|\zeta,\tau,\mu,\mathbf{k}\right\rangle\approx
\\
& -i\mu\, \frac{3\,\alpha\, t_{\perp}\left(1-3\alpha^2\right)}{\left(1+6\alpha^2\right)^2}\sin\left(\frac{\theta_{\mathbf{k}'}-\theta_{\mathbf{k}}}{2}\right)\, \left(\mathbf{k}+\mathbf{k}'\right)\cdot\mathbf{\tilde{u}}.
\nonumber
\end{align}
When the phason field is decomposed in longitudinal and transverse components, the former give rise to scalar products of the form $ i\left(\mathbf{k}+\mathbf{k}'\right)\cdot\mathbf{\hat{q}}$, while for transverse modes we have $i\left(\mathbf{k}+\mathbf{k}'\right)\cdot\left(\mathbf{\hat{z}}\times\mathbf{\hat{q}}\right)$; here $\mathbf{\hat{q}}$ is a unit vector along the transferred momentum, $\mathbf{q}=\mathbf{k}'-\mathbf{k}$. For quasi-elastic processes ($|\mathbf{k}'|=|\mathbf{k}|$) illustrated in Fig.~\ref{fig:kinematics}, we have\begin{subequations}\begin{align}
& \left(\mathbf{k}+\mathbf{k}'\right)\cdot\mathbf{\hat{q}}=0,\\
& \left(\mathbf{k}+\mathbf{k}'\right)\cdot\left(\mathbf{\hat{z}}\times\mathbf{\hat{q}}\right)=-2\left|\mathbf{k}\right|\cos\left(\frac{\theta_{\mathbf{k}'}-\theta_{\mathbf{k}}}{2}\right).
\end{align}\end{subequations}
Collecting all the pieces, I arrive at the following expression for the matrix element of the coupling with transverse phasons,\begin{widetext}\begin{align}
\varpi_{T,\tau}^{\mu}(\mathbf{q},\mathbf{k},\mathbf{k}')=-\mu\,\frac{6\,\alpha\,t_{\perp}}{1+6\alpha^2}\left(\frac{v_F^*}{v_F}\right)\left|\mathbf{k}\right|\sin\left(\frac{\theta_{\mathbf{k}'}-\theta_{\mathbf{k}}}{2}\right)\cos\left(\frac{\theta_{\mathbf{k}'}-\theta_{\mathbf{k}}}{2}\right),
\end{align}\end{widetext}
where I have regrouped some factors in $v_F^*/v_F$. This is indeed the matrix element of the coupling with $(\boldsymbol{\nabla}\times\mathbf{\tilde{u}})_z$ since for quasi-elastic processes $|\mathbf{q}|=2|\mathbf{k}|\sin(\frac{\theta_{\mathbf{k}'}-\theta_{\mathbf{k}}}{2})$. We end up then with the identification of the phenomenological parameter in Eq.~\eqref{eq:gA2}.

%\clearpage

%\section*{Supplementary material}


\begin{thebibliography}{99}

\bibitem{portu} J. M. B. Lopes dos Santos, N. M. R. Peres, and A. H. Castro Neto, Phys. Rev. Lett. \textbf{99}, 256802 (2007).

\bibitem{macdonald} R. Bistritzer and A. H. MacDonald, Proc. Natl. Acad. Sci. \textbf{108}, 12233 (2011).

\bibitem{jarillo1} Y. Cao, V. Fatemi, A. Demir, S. Fang, S. L. Tomarken, J. Y. Luo, J. D. Sanchez-Yamagishi, K. Watanabe, T. Taniguchi, E. Kaxiras, R. C. Ashoori, and P. Jarillo-Herrero, Nature \textbf{556}, 80 (2018).

\bibitem{jarillo2} Y. Cao, V. Fatemi, S. Fang, K. Watanabe, T. Taniguchi, E. Kaxiras, and P. Jarillo-Herrero, Nature \textbf{556}, 43 (2018).

\bibitem{columbia} M. Yankowitz, S. Chen, H. Polshyn, K. Watanabe, T. Taniguchi, D. Graf, A. F. Young, and C. R. Dean, Science \textbf{363}, 1059 (2019).

\bibitem{stanford} A. L. Sharpe, E. J. Fox, A. W. Barnard, J. Finney, K. Watanabe, T. Taniguchi, M. A. Kastner, and D. Goldhaber-Gordon, Science \textbf{365}, 605 (2019).

\bibitem{efetov} X. Lu, P. Stepanov, W. Yang, M. Xie, M. A. Aamir, I. Das, C. Urgell, K. Watanabe, T. Taniguchi, G. Zhang, A. Bachtold, A. H. MacDonald, D. K. Efetov, arXiv:1903.06513.

\bibitem{UCSB} M. Serlin, C. L. Tschirhart, H. Polshyn, Y. Zhang, J. Zhu, K. Watanabe, T. Taniguchi, L. Balents, and A. F. Young, arXiv:1907.00261.

\bibitem{TEM} H. Yoo, R. Engelke, S. Carr, S. Fang, K. Zhang, P. Cazeaux, S. H. Sung, R. Hovden, A. W. Tsen, T. Taniguchi, K. Watanabe, G.-C. Yi, M. Kim, M. Luskin, and E. B. Tadmor, and E. Kaxiras, and P. Kim, Nat. Mater. \textbf{18}, 448 (2019).

\bibitem{LeRoy} S. Huang, K. Kim, D. K. Efimkin, T. Lovorn, T. Taniguchi, K. Watanabe, A. H. MacDonald, E. Tutuc, and B. J. LeRoy, Phys. Rev. Lett. \textbf{121}, 037702 (2018).

\bibitem{Basov} S. S. Sunku, G. X. Ni, B. Y. Jiang, H. Yoo, A. Sternbach, A. S. McLeod, T. Stauber, L. Xiong, T. Taniguchi, K. Watanabe, P. Kim, M. M. Fogler, and D. N. Basov, Science \textbf{362}, 1153 (2018).

\bibitem{domain_walls_exp} J. S. Aldena, A. W. Tsena, P. Y. Huanga, R. Hovdena, L. Brownb, J. Parkb, D. A. Mullera, and P. L. McEuen, Proc. Natl. Acad. Sci. \textbf{110}, 11256 (2013).

\bibitem{domain_walls_th} F. Zhang, A. H. MacDonald, and E. J. Mele, Proc. Natl. Acad. Sci. \textbf{110}, 10546 (2013).

\bibitem{symmetry1} H. C. Po, L. Zou, A. Vishwanath, and T. Senthil, Phys. Rev. X \textbf{8}, 031089 (2018).

\bibitem{symmetry2} L. Zou, H. C. Po, A. Vishwanath, and T. Senthil, Phys. Rev. B \textbf{98}, 085435 (2018).

\bibitem{Zaletel} N. Bultinck, S. Chatterjee, and M. P. Zaletel, arXiv:1901.08110.

\bibitem{Zhang_etal} Y.-H. Zhang, D. Mao, and T. Senthil, arXiv:1901.08209 .

\bibitem{Xie_MacDonald} M. Xie and A. H. MacDonald, arXiv:1812.04213.

\bibitem{Paco_PNAS} F. Guinea and N. R. Walet, Proc. Natl. Acad. Sci. \textbf{115}, 13174 (2018).

\bibitem{STM1} A. Kerelsky, L. McGilly, D. M. Kennes, L. Xian, M. Yankowitz, S. Chen, K. Watanabe, T. Taniguchi, J. Hone, C. Dean, A. Rubio, and A. N. Pasupathy, Nature \textbf{572}, 95 (2019).

\bibitem{STM2} Y. Choi, J. Kemmer, Y. Peng, A. Thomson, H. Arora, R. Polski, Y. Zhang, H. Ren, J. Alicea, G. Refael, F. von Oppen, K. Watanabe, T. Taniguchi, and S. Nadj-Perge, arXiv:1901.02997.

\bibitem{STM3} Y. Jiang, J. Mao, X. Lai, K. Watanabe, T. Taniguchi, K. Haule, and E. Y. Andrei, arXiv:1904.10153.

\bibitem{jarillo3} Y. Cao, D. Chowdhury, D. Rodan-Legrain, O. Rubies-Bigord\`a, K. Watanabe, T. Taniguchi, T. Senthil, and P. Jarillo-Herrero, arXiv:1901.03710.

\bibitem{columbia_phonons} H. Polshyn, M. Yankowitz, S. Chen, Y. Zhang, K. Watanabe, T. Taniguchi, C. R. Dean, and A. F. Young, arXiv:1902.00763.

\bibitem{phonons_transport1} N. Ray, M. Fleischmann, D. Weckbecker, S. Sharma, O. Pankratov, and S. Shallcross, Phys. Rev. B \textbf{94}, 245403 (2016).

\bibitem{phonons_transport2} F. Wu, E. Hwang, and S. Das Sarma, Phys. Rev. B \textbf{99}, 165112 (2019).

\bibitem{phonons_transport3} I. Yudhistira, N. Chakraborty, G. Sharma, D. Y. H. Ho, E. Laksono, O. P. Sushkov, G. Vignale, and S. Adam, Phys. Rev. B \textbf{99}, 140302(R) (2019).

%\bibitem{phonons_MacDonald}

%\bibitem{phonons_Finland}

\bibitem{phonons_Bernevig} B. Lian, Z. Wang, and B. Andrei Bernevig, Phys. rev. Lett. \textbf{122}, 257002 (2019).

\bibitem{Mele0} E. J. Mele, Phys. Rev. B \textbf{81}, 161405(R) (2010). 

\bibitem{portu2} J. M. B. Lopes dos Santos, N. M. R. Peres, and A. H. Castro Neto, Phys. Rev. B \textbf{86}, 155449 (2012).

\bibitem{foot1} Note that $\mathbf{G}_i\cdot\mathbf{R}_j=2\pi\delta_{ij}$ by definition; equation~\eqref{eq:geometry} follows from the identity $(1-R)^T\cdot(1-R^{-1})^{-1}=1$.

\bibitem{Levine_etal} D. Levibe, T. C. Lubensky, S. Ostlund, S. Ramaswamy, P. J. Steinhardt, and J. Toner, Phys. Rev. Lett. \textbf{54}, 1520 (1985).

\bibitem{foot2} The condition of beating pattern maxima is now $\boldsymbol{\Delta}(n\,\mathbf{R}_1+m\,\mathbf{R}_2+\mathbf{\tilde{u}})=n\,\mathbf{a}_1+m\,\mathbf{a}_2-R^{-1}\mathbf{u}$; inverting this relation leads to Eq.~\eqref{eq:distances}.

%\bibitem{foot3} Comment about skyrmion lattices, helix.

%\bibitem{book}

\bibitem{rusos} A. M. Popov, I. V. Lebedeva, A. A. Knizhnik, Y. E. Lozovik, and B. V. Potapkin, Phys. Rev. B \textbf{84}, 045404 (2011).

\bibitem{Mele} X. Gong and E. J. Mele, Phys. Rev. B \textbf{89}, 121415(R) (2014).

\bibitem{Koshino} N. N. T. Nam and M. Koshino, Phys. Rev. B \textbf{96}, 075311 (2017).

\bibitem{Lame} C. Lee, X. Wei, J. W. Kysar, and J. Hone, Science \textbf{321}, 385 (2008).

\bibitem{Carr_etal} S. Carr, D. Massatt, S. B. Torrisi, P. Cazeaux, M. Luskin, and E. Kaxiras, Phys. Rev. B \textbf{98}, 224102 (2018). 

\bibitem{foot3} These one-dimensional solutions are only approximations to the domain walls between AB and BA-stacked areas, which contain both tensile and shear components in general. In the limit $L_M\ll\ell$, the latter dominates, as reflected by the strongest softening of the transverse phonon in the calculation of Fig.~\ref{fig:spectra}.

\bibitem{book} P. M. Chaikin and T. C. Lubensky, \textit{Principles of Condensed Matter Physics}, (Cambridge University Press, Cambridge, 2000).

\bibitem{foot4} If we lock the coordinate system to a beating pattern maxima then $\gamma/3=0$ modulo $2\pi$. 

\bibitem{foot_soft} As we showed before, beating pattern translations result from relative translations of one layer with respect to the other. In an incommensurate structure, this is a soft mode. Beating pattern rotations with respect to the preferential direction imposed by the superlattice potential results from expanding/compressing one layer with respect to the other, which always costs energy.

\bibitem{note_bulk} The bulk modulus $\tilde{\lambda}+\tilde{\mu}\approx\sqrt{3} V\ell/L_M$ remains positive and corresponds, roughly, to the variation of the free-energy density of the soliton network with the moir\'e period.

\bibitem{Halperin_etal} S. N. Coppersmith, D. S. Fisher, B. I. Halperin, P. A. Lee, and W. F. Brinkman, Phys. Rev. Lett. \textbf{46}, 549 (1981); Phys. Rev. B \textbf{25}, 349 (1982). 

\bibitem{fu_strain} Z. Bi, N. F. Q. Yuan, and L. Fu, Phys. Rev. B \textbf{100}, 035448 (2019).

\bibitem{Villain} J. Villain, Surf. Sci. \textbf{97}, 219 (1980).

\bibitem{Pokrovsky_Tapalov} V. L. Pokrovsky and A. L. Talapov, Phys. Rev. Lett. \textbf{42}, 65; Zh. Eksp. Teor. Fiz. \textbf{78}, 269 (1979) [Sov. Phys. JETP \textbf{51}, 134].

%{\blue \bibitem{foot_entropy} The relevance of entropic terms follows again from the special nature of the soliton network: The energy of the domain walls changes linearly with their length and there are many possible configurations in which the total length remains invariant and possesses then the same energy.}

\bibitem{phonons_2013} A. I. Cocemasov, D. L. Nika, and A. A. Balandin, Phys. Rev. B \textbf{88}, 035428 (2013).

\bibitem{Koshino2} M. Koshino, New J. Phys. \textbf{17}, 015014 (2015).

\bibitem{Koshino3} M. Koshino, N. F. Q. Yuan, T. Koretsune, M. Ochi, K. Kuroki, and L. Fu, Phys. Rev. X \textbf{8}, 031087 (2018).

\bibitem{Guinea_Walet} F. Guinea and N. R. Walet, Phys. Rev. B \textbf{99}, 205134 (2019).

\bibitem{cdw} M. F. Bishop and A. W. Overhauser, Phys. Rev. B \textbf{23}, 3638 (1981).

\bibitem{umklapp} J. R. Wallbank, R. K. Kumar, M. Holwill, Z. Wang, G. H. Auton, J. Birkbeck, A. Mishchenko, L. A. Ponomarenko, K. Watanabe, T. Taniguchi, K. S. Novoselov, I. L. Aleiner, A. K. Geim, and V. I. Fal'ko, Nat. Phys. \textbf{15}, 32 (2019). Umklapp scattering by phonons is discussed in the supplementary material appended to the electronic version of this article.

%\bibitem{foot5} Note that this IS NOT GENERAL

\bibitem{coherence1} D. Culcer and R. Winkler, Phys. Rev. B \textbf{79}, 165422 (2019).

\bibitem{coherence2} M. Trushin, J. Kailasvuori, J. Schliemann, and A. H. MacDonald, Phys. Rev. B \textbf{82}, 155308 (2010). 

\bibitem{Ziman} J. M. Ziman, \textit{Electrons and Phonons: The Theory of Transport Phenomena in Solids}, (Oxford Univesity Press, London, 1960).

\bibitem{phonons_bilayer} H. Ochoa, E. V. Castro, M. I. Katsnelson, and F. Guinea, Phys. Rev. B \textbf{83}, 235416 (2011).

\bibitem{Ando} H. Suzuura and T. Ando, Phys. Rev. B \textbf{65}, 235412 (2002).

\bibitem{Gonzalez_Stauber1} J. Gonz\'alez and T. Stauber, arXiv:1903.01376.

\bibitem{cuprates} T. M. Rice, N. J. Robinson, and A. M. Tsvelik, Phys. Rev. B \textbf{96}, 220502(R) (2017).

\bibitem{phonons_Fu} H. Isobe, N. F. Q. Yuan, and L. Fu, Phys. Rev. X \textbf{8}, 041041 (2018).

\bibitem{Gonzalez_Stauber2} J. Gonz\'alez and T. Stauber, Phys. Rev. Lett. \textbf{122}, 026801 (2019).

\bibitem{Larkin} A. I. Larkin, Zh. Eksp. Teor. Fiz. \textbf{58}, 1466 (1970) [Sov. Phys. JETP \textbf{31}, 784].

\bibitem{Katsnelson_numerics} B. Sachs, T. O. Wehling, M. I. Katsnelson, and A. I. Lichtenstein, Phys. Rev. B \textbf{84}, 195414 (2011).

\bibitem{Lee} H. Fukuyarna and P. A. Lee, Phys. Rev. B \textbf{17}, 535 (1978).

\bibitem{Larkin2} A. I. Larkin and Yu. N. Ovchinnikov, J. Low Temp. Phys. \textbf{34}, 409 (1979).

\bibitem{preprint} M. Koshino and Y.-W. Son, arXiv:1905.09660.

\bibitem{Lowdin} P.-O. L\"owdin, J. Chem. Phys. \textbf{19}, 1396 (1951).

\bibitem{foot_robustness} For this reason, the result in first-order perturbation theory is robust when the hopping amplitudes between different sublattices are changed.

\end{thebibliography}
\end{document}